\documentclass[twocol]{ametsocV6.1}

\title{The Impact of Stratification on Surface-Intensified Eastward Jets in Turbulent Gyres}

\authors{Lennard Miller\aff{a,b}, Bruno Deremble\aff{b}, Antoine Venaille \correspondingauthor{Lennard Miller, lennard.miller@ens-lyon.fr}\aff{a}}

\affiliation{\aff{a}{ENS de Lyon, CNRS, Laboratoire de Physique (UMR CNRS 5672), F-69342 Lyon, France}\\
\aff{b}{Université Grenoble Alpes, CNRS, INRAE, IRD,
Grenoble-INP, Institut des Géosciences de l’Environnement, Grenoble, France}\\
}

\abstract{This study examines the role of stratification in the formation and persistence of eastward jets (like the Gulf Stream and Kuroshio currents). Using a wind-driven, two-layer quasi-geostrophic model in a double-gyre configuration, we construct a phase diagram to classify flow regimes. The parameter space is defined by a criticality parameter \( \xi \), which controls the emergence of baroclinic instability, and the ratio of layer depths \( \delta \), which describes the surface intensification of stratification. Eastward jets detaching from the western boundary are observed when \( \delta \ll 1 \) and \( \xi \sim 1 \), representing a regime transition from a vortex-dominated western boundary current to a zonostrophic regime characterized by multiple eastward jets. Remarkably, these surface-intensified patterns emerge without considering bottom friction. The emergence of the coherent eastward jet is further addressed with complementary 1.5-layer simulations and explained through both linear stability analysis and turbulence phenomenology. In particular, we show that coherent eastward jets emerge when the western boundary layer is stable, and find that the asymmetry in the baroclinic instability of eastward and westward flows plays a central role in the persistence of eastward jets, while contributing to the disintegration of westward jets.}

\begin{document}

\maketitle

%
%
%
\statement
	 Eastward oceanic jets at mid-latitudes, such as the Gulf Stream or the Kuroshio, are important in oceanic circulation as they transport water masses from western boundary currents far into the open ocean. To isolate the role of density stratification on the formation of such jets in turbulent gyres, this study uses an idealized model with two layers of fluid stacked on top of each other. Our findings describe layer densities and depths that are favourable for jet formation, highlighting the importance of intensified flow in a thin surface layer.
%
%


\section{Introduction}

Surface-intensified eastward jets detaching from western boundaries are a prominent feature of mid-latitude oceans. Two iconic examples are the Gulf Stream and the Kuroshio, which appear as coherent, narrow, meandering ribbons surrounded by mesoscale oceanic rings in snapshots of surface kinetic energy \citep{chassignet.xu_2021}. Satellite altimetry and state-estimate reconstructions show that these jets often maintain a coherent structure over long distances, despite the transition from coastal to open-ocean environments \citep{sanchez-roman.gues.ea_2024,chassignet.marshall_2008}. Coherent eastward jets are also found across a wide range of models, from high-resolution primitive equation models \citep{uchida.le-sommer.ea_2022, ajayi.lesommer.ea_2020} to highly idealized quasi-geostrophic models \citep{holland_role_1978, berloff_large-scale_1999}. Despite their robust presence in observations and models, the understanding of the turbulent dynamics of eastward jets detaching from the western boundary remains incomplete. Here, we investigate the role of stratification in eastward jet formation using a simple two-layer quasi-geostrophic model of wind-driven ocean gyres.\\

When a western boundary current detaches it penetrates into the open ocean as an eastward jet, which tends to become unstable and shed eddies \citep{veronis1966wind, holland_role_1978}. These eddies can reinforce and sharpen the jet as it flows away from the coast \citep{wardle2000representation, greatbatch2010ocean, dritschel2011jet}. This has led numerous studies to neglect the connection of the jet to the western boundary current and instead investigate the interaction between turbulence and jets in periodic domains \citep{arbic2003coherent, arbic2004effects,gallet2021quantitative}, where jets may spontaneously emerge due to the planetary vorticity gradient \citep{rhines1975waves}. However, it remains unclear how heterogeneity and the presence of a western boundary affect the dynamics of such freely-evolving flows \citep{nadiga2010alternating, grooms_mesoscale_2013}. Statistical mechanics may explain the spontaneous emergence of eastward jets in closed domains \citep{venaille2011oceanic}, but fails to account for western intensification. Dynamical system theory can provide insights on the length of a viscous eastward jet detaching from a western boundary current \citep{simonnet2005quantization}, but is unfit to describe fully turbulent flow. To wit, there is no theoretical framework that can bridge the gap between turbulent eastward jets in the open ocean and western boundary currents.\\

In this study we aim to understand the formation of turbulent jets detaching from western boundaries by characterizing the necessary conditions for their emergence. We focus on the impact of stratification. Previous studies have shown that geostrophic turbulence can reinforce the jet in stratified models of ocean gyres \citep{sun_response_2013}, while the absence of stratification leads to its complete disintegration \citep{greatbatch2000four, fox2005reevaluating}. Two stratification properties are known to influence turbulent flow dynamics in the ocean: the depth of the pycnocline and the amplitude of the associated density jump. A shallow pycnocline favors surface-intensified dynamics \citep{fu1980nonlinear, smith.vallis_2001,meunier2023vertical}. A stronger density jump increases the internal Rossby radius of deformation, a central horizontal length scale of geostrophic turbulence \citep{vallis2017atmospheric}. For instance, this length controls turbulent regime transitions from isolated vortices to jets separating regions of homogenized potential vorticity \citep{arbic2003coherent}.\\

This article further explores the role of stratification by describing its impact on both basin-scale circulation and turbulent eastward jets. For this, we consider a two-layer quasi-geostrophic model of wind-driven ocean gyres which is the minimal model to describe both mesoscale turbulence and Sverdrup gyres. In this model, stratification is also characterized by two key parameters: the layer depth ratio, which translates to the relative depth of the pycnocline, and the internal Rossby radius of deformation. We seek to characterize transitions between flow regimes with and without jets when these two parameters are varied and interpret the results using a combination of linear stability analysis and geostrophic turbulence phenomenology.\\

The structure of the article is as follows: in section~\ref{sec:flow-model}, we describe the two-layer model and introduce the reduced parameter space which describes the vertical stratification properties. In section~\ref{sec:numerical-simulations}, we use numerical simulations to identify three distinct stratification regimes: strong, intermediate, and weak. We then shift our attention to the intermediate stratification regime. Section~\ref{sec:linear-stability-analysis} presents a local linear stability analysis of surface-intensified gyres, paying particular attention to the distinct stability characteristics of the eastward and westward parts of the gyres, as well as the western boundary layer. In section~\ref{sec:eastward-jet}, we compare a two-layer and a one-and-a-half-layer quasi-geostrophic simulation to identify the dynamics specifically linked to baroclinic instability. Finally, in section~\ref{sec:zonostrophic-regime}, we discuss the transition to a zonostrophic regime, which precedes the weak stratification regime characterized by barotropic gyres. We conclude in section~\ref{sec:conclusion} by outlining the minimal conditions required for the emergence and persistence of a coherent jet between the gyres.\\

\section{Flow Model}
\label{sec:flow-model}

\begin{table*}[h]
\caption{Parameters of the 2-layer Model for Wind-driven Oceanic Circulation.}\label{t1}
\begin{center}
\begin{tabular}{ccccrrcrc}
\topline
Name & Variable & Values in Simulations\\
\midline
Upper Layer Depth & $H_1$ & $2000$ m, $666$ m, $40$ m\\
 Lower Layer Depth & $H_2$ & $2000$ m, $3334$ m, $3960$ m\\
 Domain size & $L$ & $4000$ km\\
 Beta& $\beta$ & $1.7\times10^{-11}$ m$^{-1}$ s$^{-1}$\\
 Viscosity& $\nu$& $10$ m$^2$ s$^{-1}$\\
 Deformation Radius & $L_d$ & $2.3$ km, ... , $720$ km\\
 Wind Stress& $\tau_0$& $50 \times 10^{-5}$ m$^2$ s$^{-2}$, $16 \times 10^{-5}$ m$^2$ s$^{-2}$, $1 \times 10^{-5}$ m$^2$ s$^{-2}$\\
 Sverdrup Velocity scale& $U_{Sv}$& $0.09$ m$^2$ s$^{-1}$\\
\botline
\end{tabular}
\end{center}
\end{table*}
The simplest model to address the effect of stratification on wind-driven circulation consists of two layers of fluid following quasi-geostrophic motion, considered here in a double-gyre configuration. The equations of evolution of the upper layer and lower layer potential vorticity ($q_1$ and $q_2$) are

\begin{align}
&\frac{\partial q_1}{\partial t} + \mathbf{u}_1 \cdot \nabla q_1 = \frac{\nabla \times \boldsymbol{\tau} }{H_1}  + \nu \nabla^4 \psi_1 
\label{dynamical_system1}\\
&\frac{\partial q_2}{\partial t} + \mathbf{u}_2 \cdot \nabla q_2 = \nu \nabla^4 \psi_2 
\label{dynamical_system2}\\
&q_1 = \nabla^2 \psi_1 + \frac{1 - \delta}{L_d^2} (\psi_2 - \psi_1) + \beta y \\
&q_2 = \nabla^2 \psi_2 + \frac{\delta}{L_d^2} (\psi_1 - \psi_2) + \beta y
\end{align}
with $\psi_i$ the streamfunction and $\mathbf{u}_i = (u_i, v_i) = (-\partial_y \psi_i, \partial_x \psi_i)$ the velocity in each layer. We solve this system of equations on a square domain with $(x, y) \in [0, L]^2$. Since our study focuses primarily on regimes with surface-intensified turbulent flow, we consistently neglect linear drag which is often thought to represent bottom friction in the bulk of the domain. We instead choose to dissipate energy along the domain boundary by applying lateral no-slip boundary conditions \citep{miller_gyre_2024}.\\

The forcing $\boldsymbol{\tau} = (\tau_0\sin(2\pi y/L),0)$ models a classical double gyre. Stratification is completely described by the relative upper layer thickness and the first baroclinic Rossby radius of deformation, defined respectively as
\begin{equation}
\delta = \frac{H_1}{H_1 + H_2}, \quad L_d = \sqrt{\frac{g' H_1 H_2}{(H_1 + H_2) f_0^2}},
\end{equation}
with $H_i$ being the layer thicknesses, $g'$ the reduced gravity between both layers and $f_0$ the Coriolis parameter. In addition to $\delta$, the problem admits three non-dimensional parameters:
\begin{equation}
\tilde{\nu} = \frac{\nu}{U_{Sv} L}, \quad \tilde{\beta} = \frac{\beta L^2}{U_{Sv}}, \quad \xi = \frac{U_{Sv}}{\beta L_d^2}
\end{equation}
with
\begin{equation}
    U_{Sv} =\frac{8\pi \tau_0} {H_1 L \beta}
\end{equation} 
being the velocity scale of the Sverdrup flow. For all simulations, we set $L = 4000$~km, $\beta = 1.7\times10^{-11}$~m$^{-1}$\,s$^{-1}$, $\nu = 10$~m$^2$\,s$^{-1}$, and $U_{Sv}= 0.09$~m\,s$^{-1}$. The parameters $\tilde{\nu}$ and $\tilde{\beta}$ are thus kept constant at $\tilde{\beta} \simeq 3000$, $\tilde{\nu}\simeq 10^{-5}$. This corresponds to a thickness of the inertial boundary layer $\delta_I = L / \sqrt{\tilde{\beta}} = 72$~km \citep{charney1955gulf} and a viscous sublayer of $\delta_P = L \sqrt{\tilde{\nu}/\sqrt{\tilde{\beta}}} = 3$ km \citep{pedlosky1987geophysical, ierley_analytic_1986}.\\

In this study, we focus on the dynamical regimes that appear upon varying the two parameters $\xi$ and $\delta$. $\delta$ is a non-dimensional measure of surface-intensification of the stratification. In two-layer models of planetary gyres, $\delta$ may be interpreted as the position of the pycnocline and is often placed at $\delta = 0.1 - 0.2$ \citep{flierl_1978}. However, as shown in \cite{smith.vallis_2001}, the scenario $\delta \ll 1$ is an interesting asymptotic limit because it corresponds to a situation where barotropization is halted at the first baroclinic mode, contrary to the classical picture where energy cascades to the barotropic mode \citep{salmon1998lectures}. In order to investigate this limit, we will consider the range $0.01 \le \delta\le 0.5$ and focus mostly on $\delta = 0.01$. Although this value is probably beyond the realistic value observed in the ocean, the dynamical regimes were easier to entangle for such a small value of $\delta$.\\

The parameter $\xi$ is a non-dimensional deformation radius sometimes called the criticality parameter, related to baroclinic instability of the two-layer model \citep{stone_1978}. We recall here the interpretation of \cite{held1996scaling} and \cite{jansen.ferrari_2012} adapted to our configuration. Suppose there is a zonal Sverdrup flow of magnitude $U_{Sv}$ confined to the upper layer. The associated potential vorticity gradient in the upper and lower layer is
 
 \begin{equation}
     \frac{\partial Q_1}{\partial y} = \beta + \frac{(1 - \delta) U_{Sv}}{L_d^2}\,, \quad \textrm{and}  \quad \frac{\partial Q_2}{\partial y} = \beta - \frac{\delta U_{Sv}}{L_d^2}\, .
 \end{equation}
 
For baroclinic instability to occur, these gradients have to be of opposite sign \citep{pedlosky1987geophysical}. For westward flow ($U_{Sv}<0$) the gradient of the lower layer is always positive, and the condition for baroclinic instability is reached when  $\partial Q_1/\partial y < 0$. In the limit $\delta \ll 1$, this corresponds to $\xi >1$. For eastward flow ($U_{Sv}>0$) the gradient of the upper layer is always positive and the condition for instability is given by $\partial Q_2/\partial y < 0$, which corresponds to $\delta \xi >1$. $\xi$ therefore serves as a proxy for the baroclinic stability of the large-scale flow, where $\xi < 1$ corresponds to stable flows. Furthermore, there is a second interpretation of $\xi$ in ocean gyres because we can write
\begin{equation}
    \xi = \frac{\delta_I^2}{L_d^2}\, ,
\end{equation}
Then, $\xi$ is a non-dimensional measure of the width of the inertial boundary layer relative to the deformation radius.\\

In order to reach asymptotic behaviour, we vary $\xi$ from $10^{-2}$ to $10^2$ for $\delta = 0.5$, from $1$ to $5.6$ for $\delta = 1/6$ and from $10^{-1}$ to $10^3$ for $\delta = 0.01$. This corresponds to changing $L_d$ from $737$ km to $7$ km, $72$ km to $13$ km and $233$ km to $2$ km, respectively. In our numerical simulations, we change $L_d$ by setting $g'$, which controls $\xi$ but leaves all other non-dimensional variables unchanged. A summary of the dimensional variables of the simulations is given in table~\ref{t1}.\\

To numerically solve equations (\ref{dynamical_system1}) - (\ref{dynamical_system2}) the code \texttt{qgw} was used. It uses classical finite differences discretization of the equations with an Arakawa formulation of the Jacobian and a direct spectral inversion (after projection on the vertical modes) to obtain the streamfunction $\psi$ from the potential vorticity $q$. Time stepping is done with an adaptive 2nd order Adams-Bashforth technique, where the time step is determined by a standard CFL-condition. Unless specified, all runs are performed with 4096$\times$4096 grid points on the horizontal, corresponding to a resolution of $1$ km, ensuring all relevant flow scales of the problem are well resolved. We paid attention to run the simulations for at least two gyre turnover times $T_{Gyre} = L/U_{Sv}$ after the spinup was completed. Last, the no-slip boundary condition is implemented as in \cite{miller_gyre_2024}.

\section{Numerical Simulations}
\label{sec:numerical-simulations}

\subsection{Reference Run}

\begin{figure*}[!hbt]
    \centering
    \includegraphics[width = 0.8\textwidth]{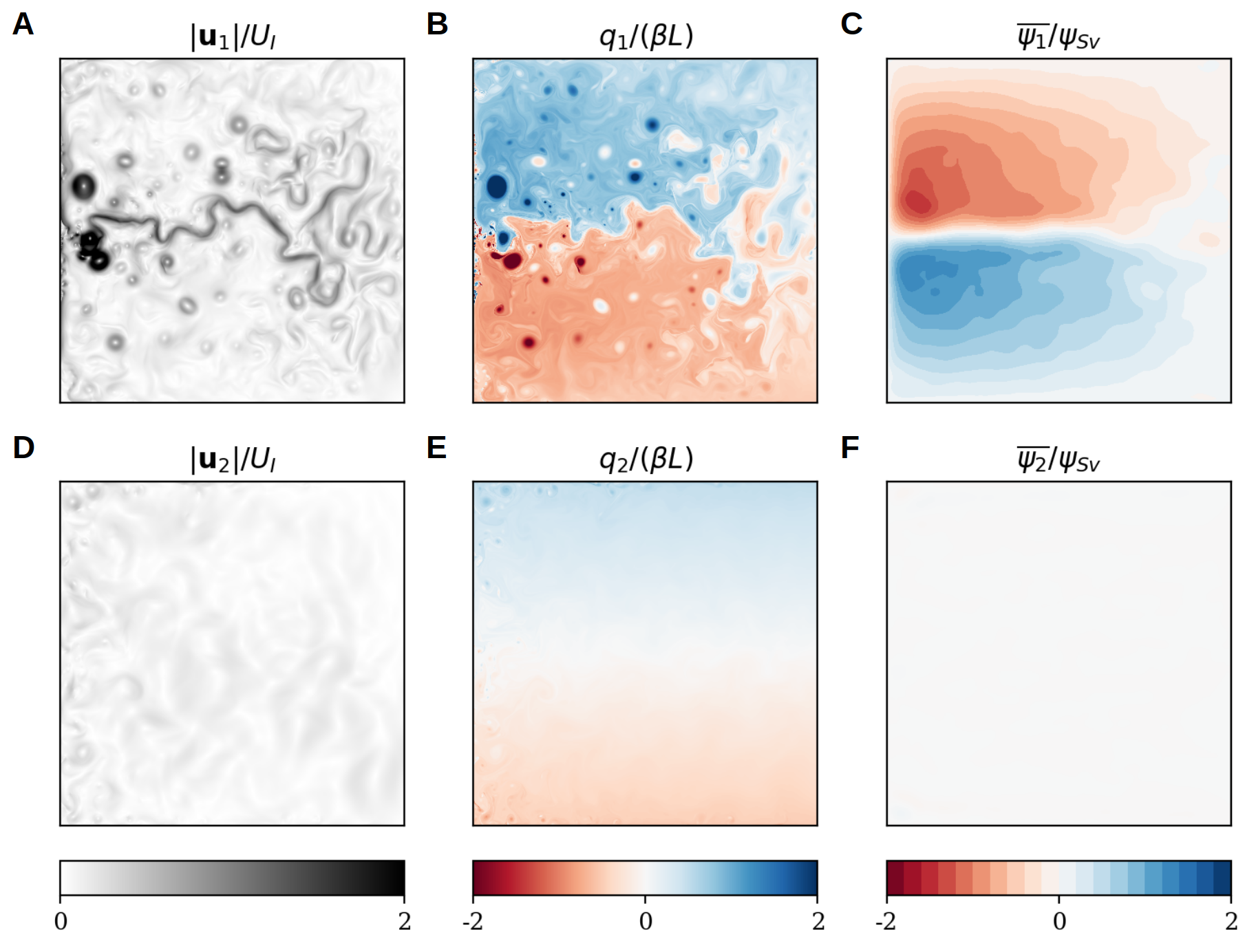}
    \caption{Upper layer (top row) and lower layer (bottom row) for the reference run at $\delta = 0.01$, $\xi = 3.3$. (\textbf{A}, \textbf{D}) Flow speed $|\mathbf{u}_i|$ normalized with inertial velocity $U_I=\psi_{Sv}/\delta_I$, (\textbf{B}, \textbf{E}) potential vorticity normalised by planetary vorticity and (\textbf{C}, \textbf{F}) time-mean streamfunction normalised by Sverdrup scaling.}
    \label{fig: ref_run}
\end{figure*}

A reference run displaying an eastward jet detaching from the western boundary is shown in figure $\ref{fig: ref_run}$, where the barred quantities denote time-averages. The criticality and layer aspect ratio are set to $(\xi, \delta) = (3.3, 0.01)$, corresponding to a deformation radius $L_d = 40$~km. The flow is western- and surface-intensified in both instantaneous and time-averaged fields, and total transport scales well with the Sverdrup scaling $\psi_{Sv} = U_{Sv}L/4$. The upper-layer potential vorticity is nearly homogenized around values of $ \psi_{Sv} / L_d^2$ in each gyre, with the potential vorticity jump at the interface between the two gyres corresponding to an eastward jet. In instantaneous kinetic energy snapshots, this jet appears with a thickness close to $L_d$ and sheds blobs of potential vorticity that have opposite signs of potential vorticity to the background, similar to Gulf stream rings  \citep{kurashina_western_2021}). Our aim in this article is to explain how and when stratification renders such states possible in double-gyre configurations without bottom friction.\\

\begin{figure*}
    \centering
    \includegraphics[width = \textwidth]{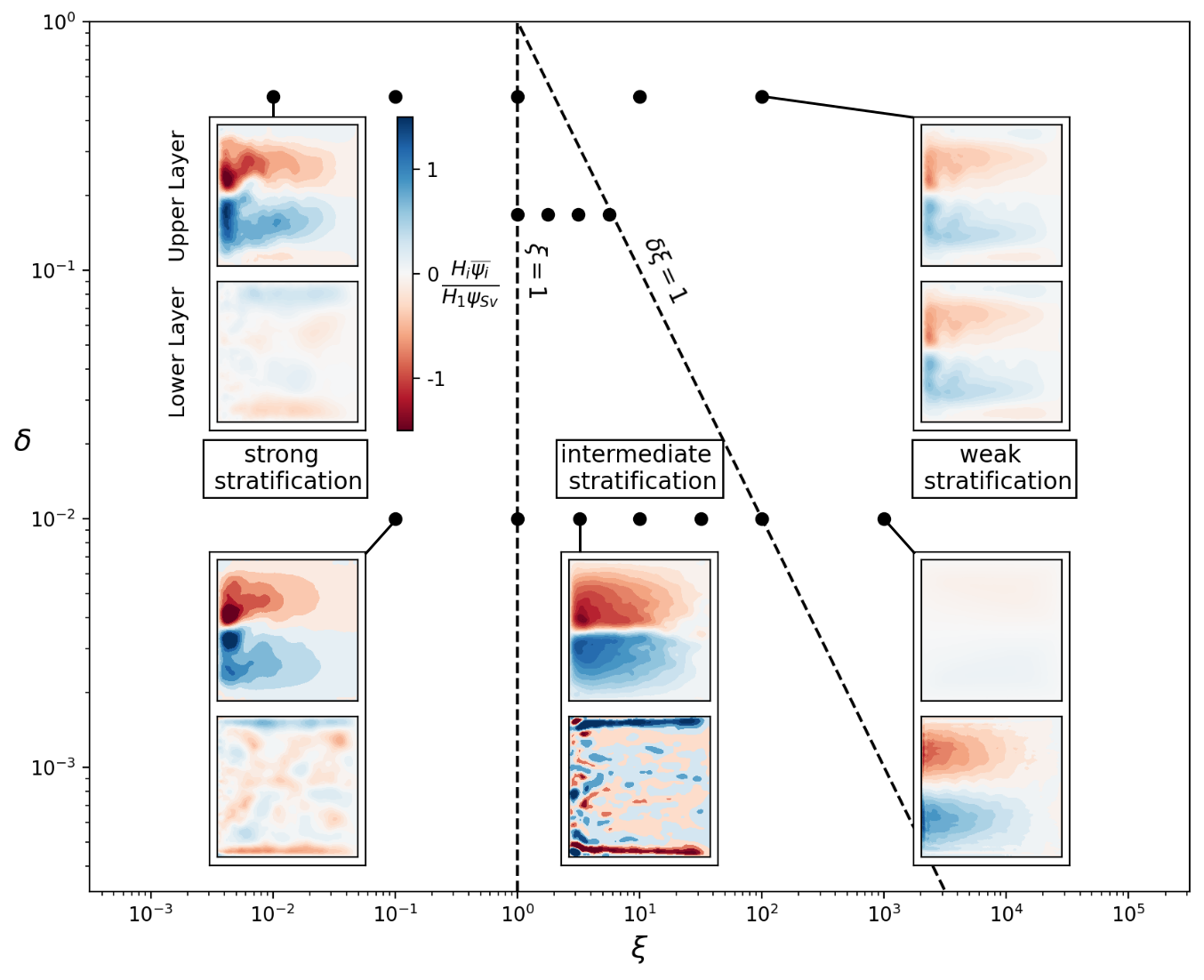}
    \caption{Parameter space spanned by $\xi$ and $\delta$. The vertical line at $\xi = 1$ marks the onset of baroclinic instability, while the slanted line $\delta\xi = 1$ signifies the barotropization of a Sverdrup flow. Black dots mark simulation parameters. The inlets show the time-mean vertically integrated transport in the upper and lower layer for example runs of each regime. The colorscale is normalized by the Sverdrup transport $\psi_{Sv} = U_{Sv}L/4$.}
    \label{fig: param_space}
\end{figure*}

\subsection{Parameter Space}

\begin{figure*}[!htb]
    \centering
    \includegraphics[width=\linewidth]{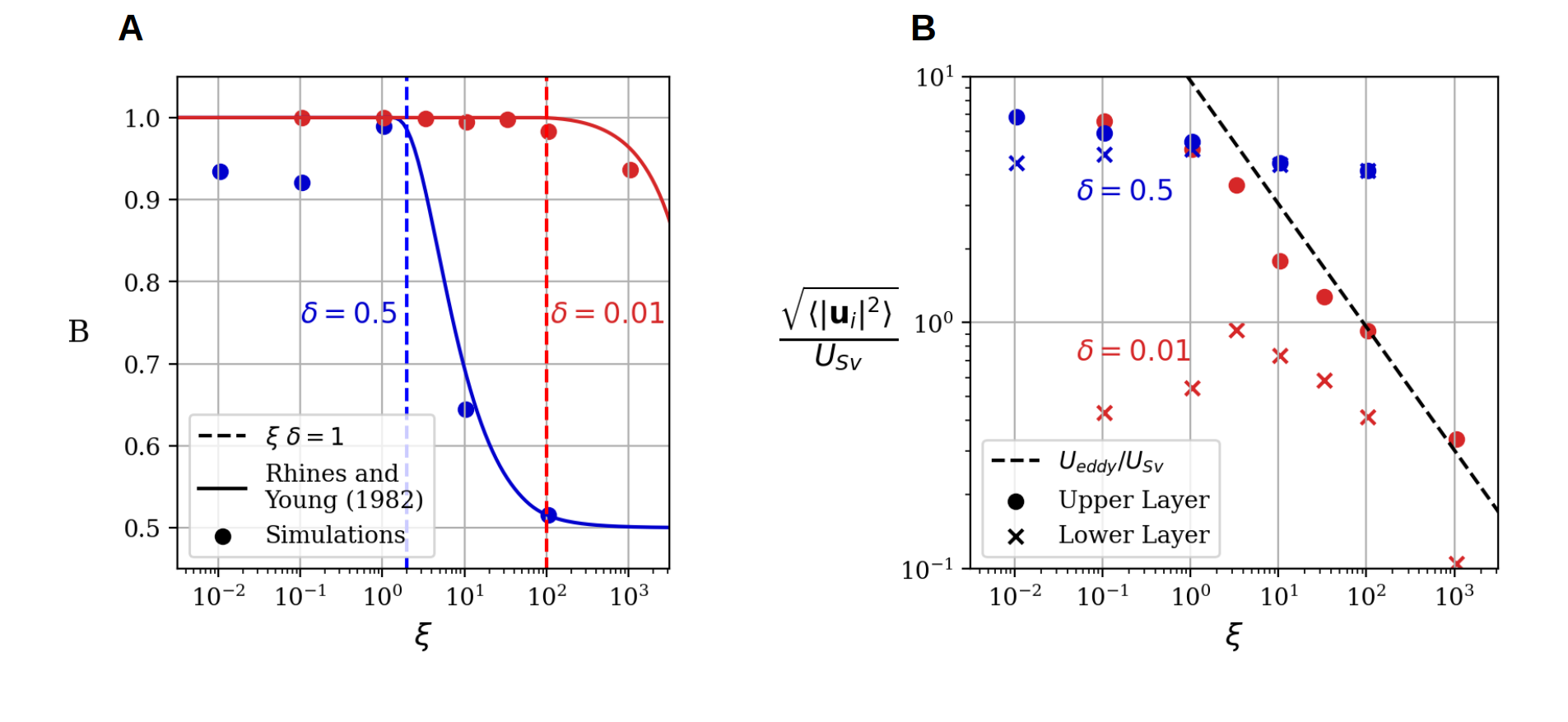}
    \caption{(\textbf{A}) Barotropization measure $B$ of the mean Sverdrup flow as defined in equation (\ref{eq: barotropization}), and (\textbf{B}) root mean square velocity as a function of $\xi$. For details on the predictions of homogenization theory from \cite{rhines_homogenization_1982} in \textbf{A} the reader is referred to appendix~B.}
    \label{fig: global diags}
\end{figure*}

In order to understand the robustness of eastward jets detaching from western boundaries we now explore a broader stratification parameter range by spanning the  parameter space $(\xi,\delta)$, focusing on the asymptotic values of $\delta = 0.5$ and $\delta = 0.01$. This parameter space, displayed in figure~\ref{fig: param_space}, is dissected into a strong stratification regime when $\xi < 1$, a weak stratification regime when $1/\delta < \xi$, and an intermediate stratification regime in between. As detailed in the next section, the transitions at $\xi=1$ and $\xi=1/\delta $ correspond to the onset of baroclinic instability for westward flow and eastward flows, respectively.
The line $\xi =1/\delta$ also corresponds to the onset of  barotropization of the Sverdrup gyres, according to classical theory of potential vorticity homogenization in planetary gyres (\cite{rhines_homogenization_1982}, see also appendix~B). To measure surface intensification of the large-scale velocity field we compute the ratio
\begin{equation}
    B = \frac{|u_1^{proj}|}{|u_1^{proj}| + | u_2^{proj}|}
    \label{eq: barotropization}
\end{equation}
with
\begin{equation}
    u_i^{proj} = \frac{2}{L^2}\iint \overline{u_i} \ \cos\left(\frac{2\pi y}{L} \right) dx \ dy,
\end{equation}
shown in figure~\ref{fig: global diags}. The projection on the gravest meridional mode is used to remove rectified zonal structures \citep{fox2004wind2, nadiga2010alternating}, and we use the time-averaged zonal velocity $\overline{u_i}$ in order to weaken the imprint of western boundary currents not accounted for in classical theory \citep{rhines_homogenization_1982}. $B$ decreases from $1$, corresponding to a complete surface intensification,  to $0.5$, corresponding to an equality between surface and lower layer velocities, which we define here as the barotropic state. \\

In the strong stratification regime ($\xi < 1$),  gyres remain confined to the upper layer, but Gulf Stream-like jets are not observed. Instead, instantaneous fields display strong vortices generated through the detachment of the viscous sublayer at the western boundary (not shown). This regime corresponds to Gyre Turbulence as reported on in \cite{miller_gyre_2024}, except that the finite value of $L_d$ in the present case tends to stabilize an intense vortex dipole close to the western boundary. The time-averaged bulk structures are similar for both values of $\delta$ and resembles a Sverdrup gyre, albeit with notable rectification (figure~\ref{fig: param_space}, left panels). Although we filter the signal, these rectified barotropic gyres at $\delta = 0.5$ which form close to the meridional boundaries correspond to $B$ less than $1$ due to their projection on the gravest meridional mode (figure~\ref{fig: global diags}A), but in fact all Sverdrup transport remains confined in the upper layer. We attribute the rectification to the presence of barotropic basin modes which can energize the lower layer when $\delta \sim 1$ (figure~\ref{fig: global diags}B), but expect a complete decoupling between the two layers in the limit of an infinite Rossby radius of deformation (corresponding to $\xi \rightarrow 0$) regardless of the value of $\delta$.\\ 

In the weak stratification regime ($\xi > 1/\delta$) the gyres become increasingly barotropic, as measured by $B$ (figure~\ref{fig: global diags}A). The flow at $\delta = 0.5$ closely resembles the barotropic Gyre Turbulence regime observed by \cite{miller_gyre_2024}. This barotropization is visible in the collapse of upper and lower layer kinetic energies and in the decrease of the index $B$. For $\delta = 0.01$ the flow enters the weak stratification regime when $\xi > 100$, and the transition is less abrupt than for $\delta = 0.5$. The time-averaged barotropic gyres align more closely with Sverdrup balance, visible in the decline of $B$ (also in the bottom right panel of figure~\ref{fig: param_space}). We find large areas of homogenized potential vorticity in the lower layer for both $\delta = 0.5$ and $\delta = 0.01$. However, the upper layer flow is no longer in Sverdrup balance but instead shows a loss of western intensification. For further details on the weak stratification regime and potential vorticity homogenization the reader is referred to appendix~B.\\

\begin{figure*}
    \centering
    \includegraphics[width = 0.8\textwidth]{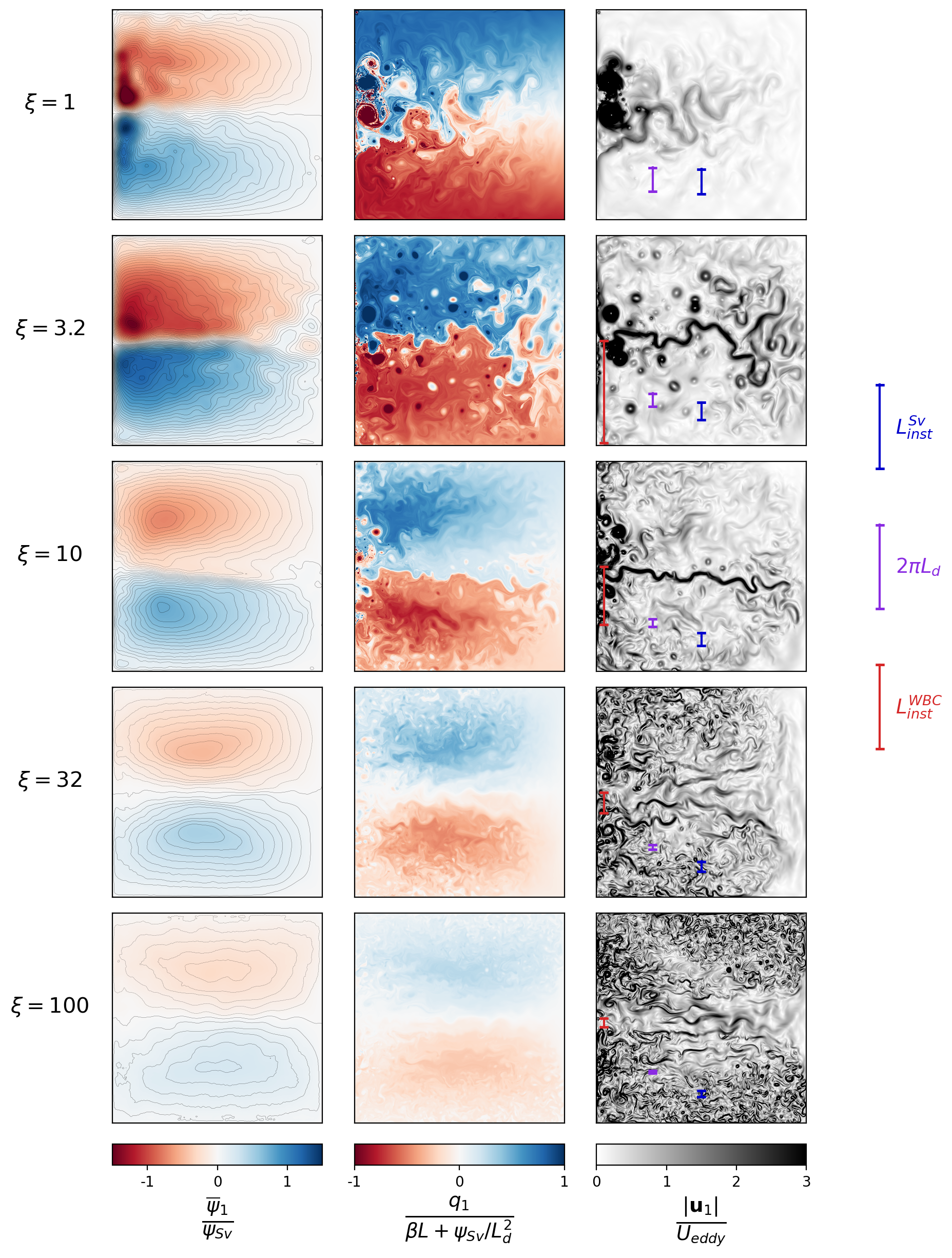}
    \caption{Time-average of $\psi_1$ (left column) and snapshots of $q_1$ (middle column) and $|\textbf{u}_1|$ (right column) for simulations in the intermediate stratification regime at $\delta = 0.01$. The stream functions are normalized by the Sverdrup scale $\psi_{Sv}$ and contour intervals are $0.05$. $q_1$ is normalized by the planetary geostrophic vorticity of Sverdrup flow ($\beta L + \psi_{Sv}/L_d^2$) and the velocity scale $U_{eddy}$ is given in equation (\ref{eq: U_turb}). The brackets shown on the right allow to compare the instability scales of the western boundary current $L_{inst}^{WBC}$ and of the interior Sverdrup flow $L_{inst}^{Sv}$ to $L_d$.}
    \label{fig: jet_transition}
\end{figure*}

The focus of this manuscript lies on the intermediate stratification regime, due to the appearance of eastward jets in gyres such as presented in the reference run. This regime is characterised by $1 < \xi < 1/\delta$ and is hence accessible only when $\delta \ll 1$. We therefore focus on the runs at $\delta = 0.01$, limiting our description to the upper layer as the dynamics remain mostly surface-intensified (figure~\ref{fig: global diags}). The surface flow patterns are displayed in figure~\ref{fig: jet_transition}. As $\xi$ is increased, a continuous transition occurs from western-intensified gyres to a state that restores east-west symmetry, with a decrease in total transport of the gyres. Coincidentally, the dominant turbulent flow features change. The strong coherent vortices which dominate the flow in the strong stratification regime disappear, and instead a sea of eddies of characteristic size $L_d$ develops in the regions of westward flow. In the regions of eastward flow, the transition is marked by the emergence of strong jets, initially as a single jet ($1 \lesssim \xi \lesssim 10$) and later as multiple jets associated with potential vorticity staircases ($10 \lesssim \xi$).  \\

The runs which were performed at $\delta = 1/6$ are omitted in the main text as no eastward jet was observed in this case. Its absence is interpreted as a result of the energetic vortex gas that forms at low viscosity when no-slip boundary conditions are applied \citep{miller_gyre_2024}. Further details on these runs can be found in appendix~A. \\ 
 
In the remainder of this paper, we use a combination of linear stability analysis, scaling analysis and complementary simulations to rationalize the transition occurring in the intermediate stratification regime at $\delta = 0.01$ (figure~\ref{fig: jet_transition}), rationalizing both the emergence and disappearance of the eastward jet detaching from the western boundary.\\

\section{Linear Stability Analysis}
\label{sec:linear-stability-analysis}

Here, we provide a linear stability analysis of the mean flow associated with surface-intensified gyres and inertial western boundary layers. This analysis will help interpreting the parameter space laid out in the previous section. To simplify, we consider two sub-problems:\\

\begin{itemize}
\item A horizontally homogeneous eastward or westward flow confined to the upper layer, with prescribed magnitude $U_{Sv}$.
\item A meridional velocity profile typical of inertial western boundary currents. This velocity profile varies in the $x$-direction over a scale of $\delta_I$ and is invariant in the y-direction.
\end{itemize}
The first case is the textbook Phillips problem for baroclinic instability on the beta plane \citep{vallis2017atmospheric}. Similar cases to the second one can be found in \cite{ierley1991viscous} and \cite{berloff_quasigeostrophic_1999}. Here, we present only the key results relevant for interpreting the nonlinear simulations in the parameter space $(\xi, \delta)$. For more details, the reader may consult appendix~C.\\

\begin{figure*}[!htb]
    \centering
    \includegraphics[width = 0.9\textwidth]{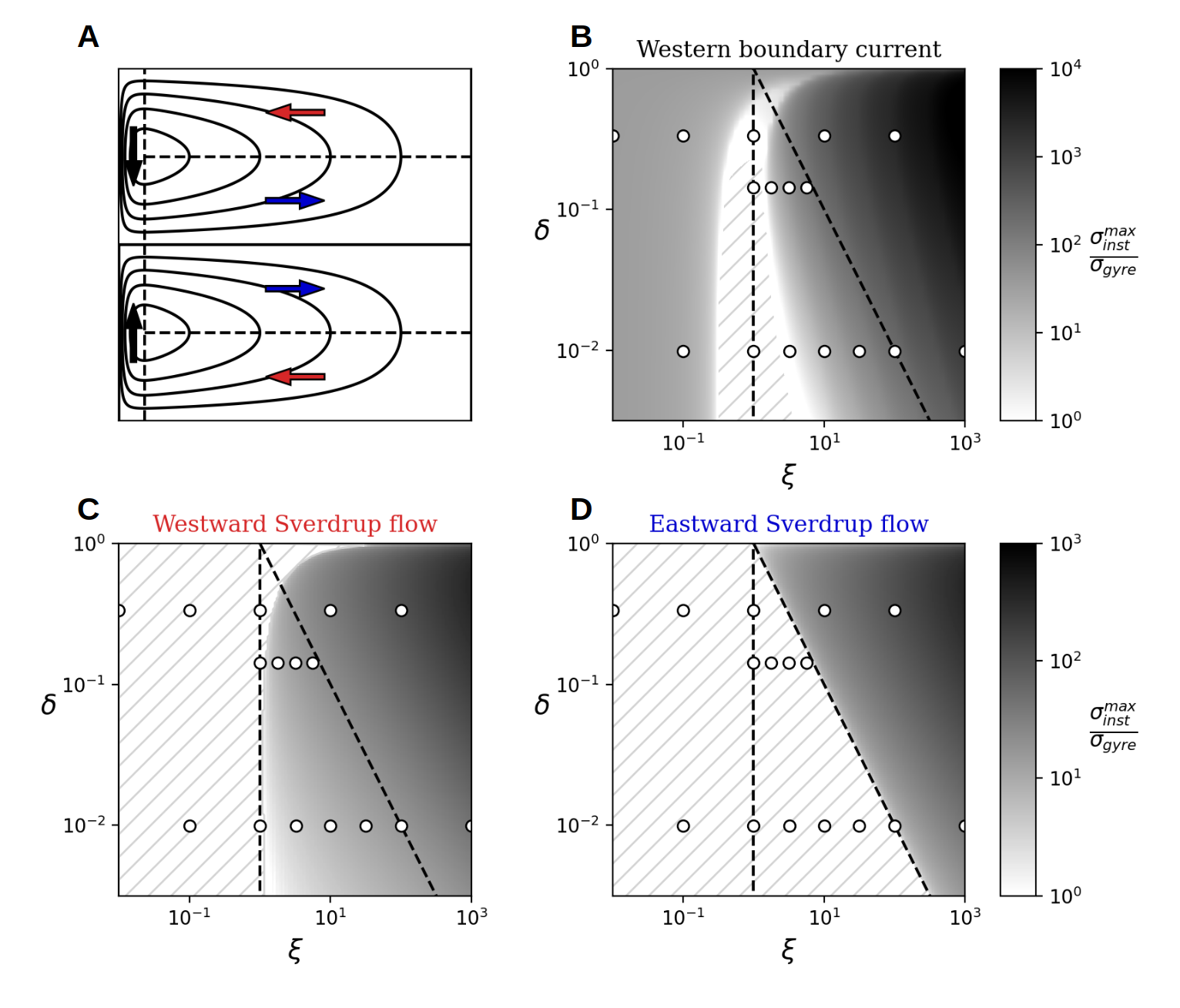}
    \caption{Results of linear stability analysis of interior Sverdrup flow with an inertial western boundary current. (\textbf{A}) Schematic of the reprented regions, and maximum growth rates for (\textbf{B}) western boundary currents, (\textbf{C}) westward Sverdrup flow and (\textbf{D}) eastward Sverdrup flow. The system is stable in the hatched regions. In the unstable domain, the growth rate (black shading) is rescaled by the gyre turnover rate, $\sigma_{gyre} = U_{Sv}/L$, and the colorbar for eastward flow applies for westward flow, too.}
    \label{fig: instability}
\end{figure*}

Baroclinic instability in the Phillips model on the beta-plane is known to be asymmetric with respect to the zonal direction of the mean flow \citep{pedlosky1987geophysical}. This asymmetry is illustrated in figure~\ref{fig: instability}, which shows the maximum growth rate as a function of $\delta$ and $\xi$ for both eastward and westward Sverdrup flows. The asymmetry appears in the intermediate stratification regime: westward Sverdrup flow becomes unstable for $\xi>1$, while eastward Sverdrup flow becomes unstable for $\xi>1/\delta$. In the weak stratification regime, where surface-intensified flow in both directions is unstable, Sverdrup flow becomes barotropic according to potential vorticity homogenization theory, thus invalidating our initial hypothesis of a surface-intensified gyre.\\

\begin{figure*}
    \centering
    \includegraphics[width=0.8\linewidth]{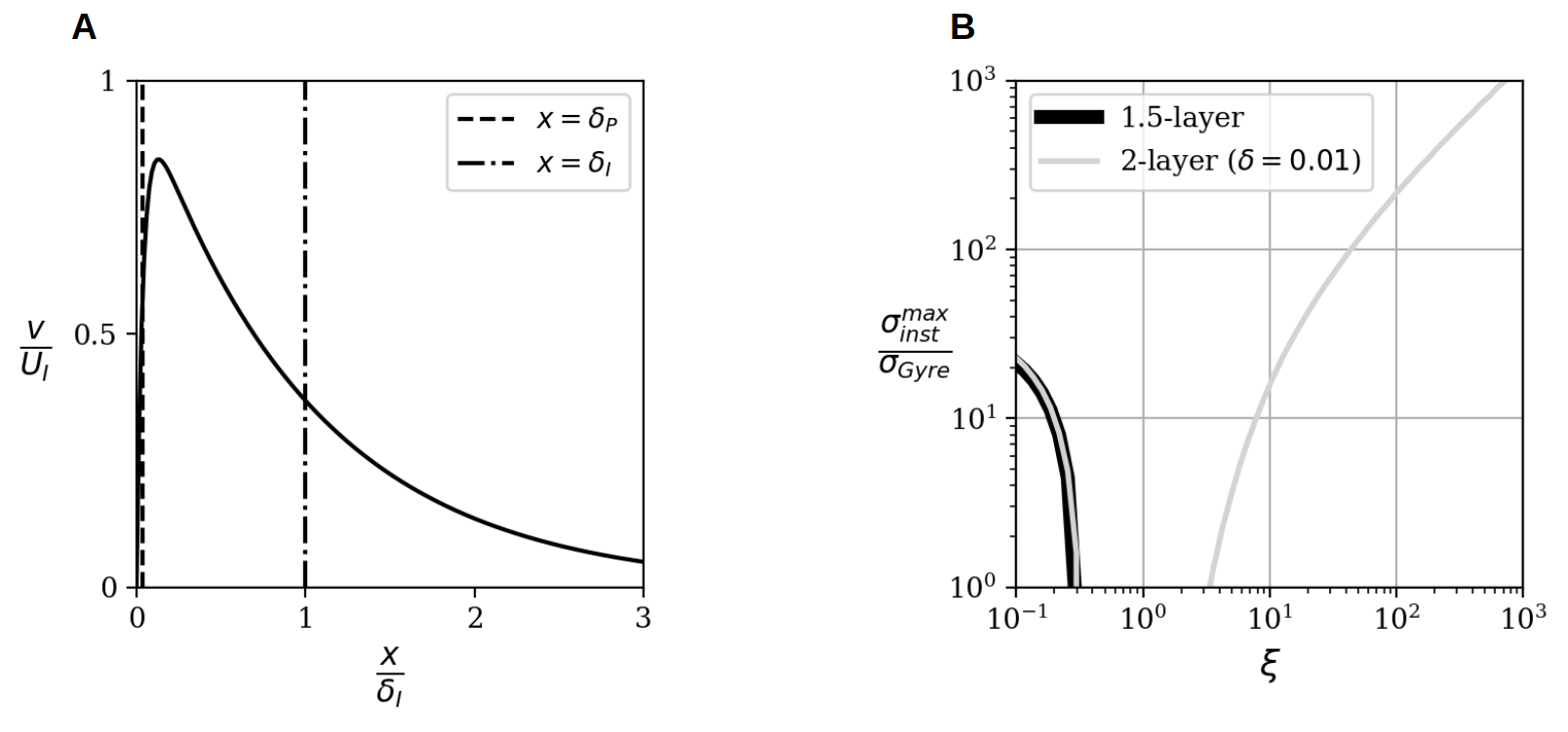}
    \caption{(\textbf{A}) Base velocity profile for the linear stability analysis of the inertial western boundary layer and (\textbf{B}) comparison of growth rates in the 1.5-layer model and the 2-layer model. The unstable branch at $\xi > 1$ visible in the 2-layer model only is attributed to baroclinic instability.}
    \label{fig: western instability}
\end{figure*}

We also conducted a linear instability analysis of the western boundary layer, as shown in figure  \ref{fig: western instability}. The base profile of the meridional velocity, depicted in figure~\ref{fig: western instability}A, is assumed to be

 \begin{equation}
     v = U_I(e^{-\frac{x}{\delta_I}} - e^{-\frac{x}{\delta_P}}).
     \label{eq: base profile}
 \end{equation}
It follows an exponential function with a decay rate determined by the inertial layer thickness $\delta_I$. Additionally, a viscous sublayer is modeled by including another exponential term with a decay rate of $\delta_P$, bringing the velocity to zero near the western boundary. The velocity amplitude is determined by the inertial velocity $U_I = \psi_{Sv}/\delta_I$, ensuring that the gyre's transport fully recirculates within this boundary layer profile. Further details can be found in Appendix~C. The growth rate of the most unstable mode in this case is plotted as a function of $(\xi,\delta)$ in figure~\ref{fig: instability}B and in figure~\ref{fig: western instability}B for the case $\delta=0.01$. An important result of this analysis is the existence of a stability island around $\xi = 1$ for sufficiently small $\delta$. Near $\xi = 1$, we observe that a meridional flow with characteristic width $\delta_I \simeq L_d$ is stable. This adds to the work of \cite{spall_generation_2000} who showed that a uniform meridional flow is necessarily baroclinically unstable. In figure~\ref{fig: western instability}B, we also plot the growth rate of the most unstable mode for the same velocity profile in a 1.5-layer quasi-geostrophic model, obtained by assuming $\psi_2 = 0$. This demonstrates that instabilities at $\xi < 1$ are due to horizontal shear instabilities of the boundary layer, while instabilities at larger criticality parameters are genuinely baroclinic, as they are absent in the 1.5-layer model.\\

In figure~\ref{fig: jet_transition} (right panels), we plot the wavelength of the most unstable mode for both the western meridional flow and the 
zonal Sverdrup flow, to compare this length scale with flow structures. For the Sverdrup flow, the most unstable mode has a length scale $2\pi L_d$, which appears to qualitatively match the size of the vortex rings. In contrast, for the western boundary current, the instability scale is initially much larger than the deformation radius, and only at values of $\xi > 10$ approaches the order of the deformation radius. Snapshots of $\psi$ show the presence of large-scale basin modes (not shown), which may result from this instability at intermediate $\xi \sim 10$. At higher $\xi \sim 100$, baroclinic instability is supposedly of the same type as in the Sverdrup flow, as the instability scale becomes much smaller than the scale of the mean profile, $\delta_I$. However, at this point the mean flow is no longer western intensified (figure~\ref{fig: jet_transition}A). Baroclinic instability itself may in fact be responsible for this erosion of Sverdrup flow, but a detailed analysis of the mean flow balance is beyond the scope of the present work.\\

\section{The Eastward Jet ($1 \lesssim \xi \lesssim 10$): Insights from the 1.5-Layer Model}
\label{sec:eastward-jet}

\begin{figure*}[!htb]
    \centering
    \includegraphics[width = \textwidth]{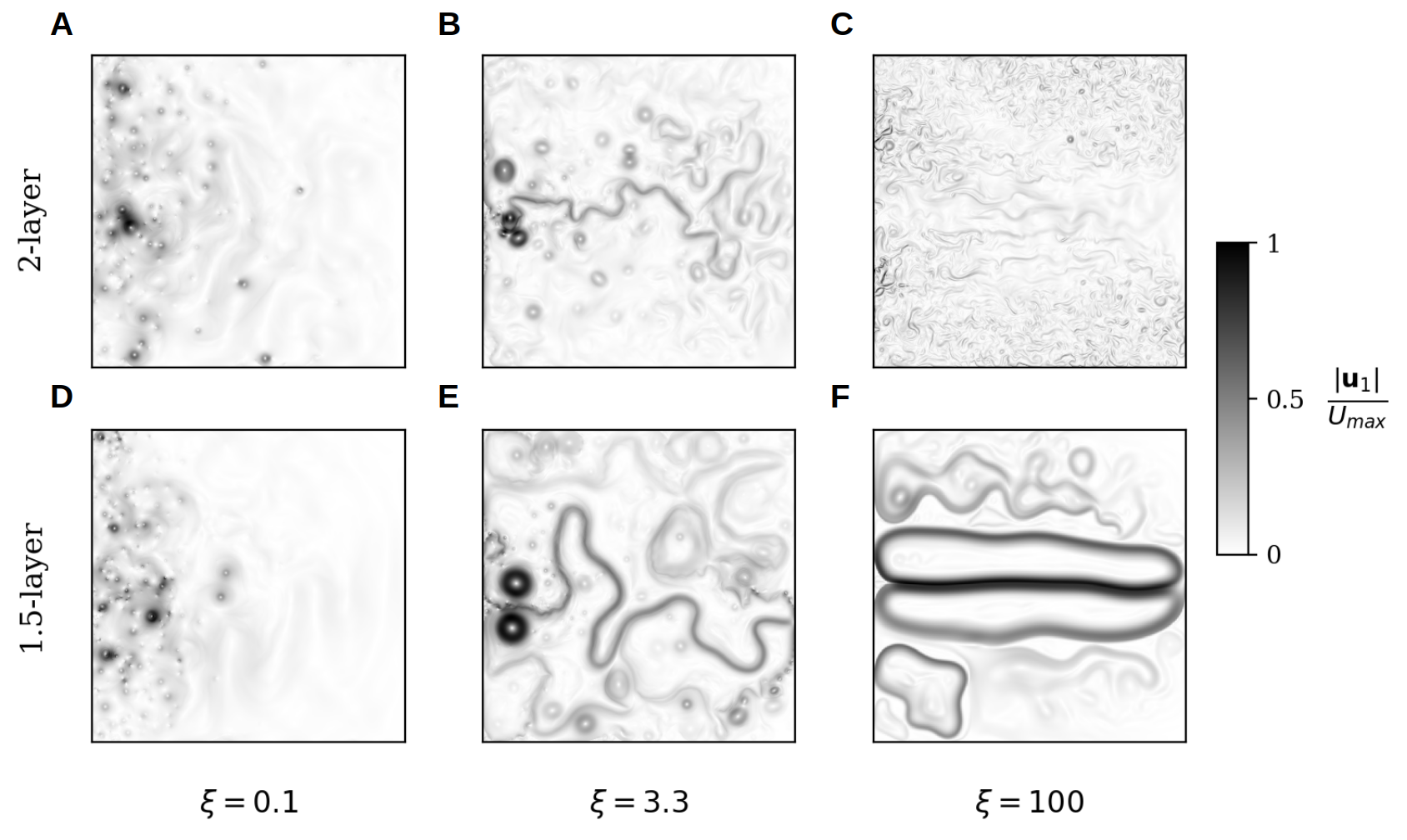}
    \caption{Comparison of instantaneous surface velocity snapshots between the 2-layer model at $\delta = 0.01$ (top row) and the 1.5-layer model (bottom row), at (\textbf{A},\textbf{D}) $\xi = 0.1$, (\textbf{B},\textbf{E}) $\xi = 3.3$ and (\textbf{C},\textbf{F}) $\xi = 100$.  The maximum velocities in the instantaneous flow fields (renormalized by $U_{Sv}$) for the 2-layer model are $372, 166$ and $50$, for the 1.5-layer model they are $361, 208$ and $105$. The simulation at $\xi = 100$ in the 1.5-layer model is still in the process of spin-up.}
    \label{fig: 1.5-layer}
\end{figure*}

To disentangle the role of baroclinic instability and intrinsic surface layer dynamics in the emergence of the eastward jet we performed a set of 1.5-layer numerical simulations, assuming $\psi_2=0$ at all times in equation (\ref{dynamical_system2}) with all other parameters kept identical. A comparison between the two-layer model and the 1.5-layer model is shown with plots of the instantaneous speed in figure~\ref{fig: 1.5-layer}.\\

In the strong stratification regime ($\xi<1$), both models yield a similar flow pattern, resembling the Gyre Turbulence regime of \cite{miller_gyre_2024}. This suggests that the dynamics are primarily governed by 1.5-layer dynamics, consistent with the linear computation predicting no baroclinic instability in this parameter range. As $\xi$ becomes larger than one, both models simulate a change of regime with the emergence of strong eastward jets. However, a notable difference occurs: recirculating westward jets appear in the 1.5-layer model but not in the two-layer model (see figure~\ref{fig: 1.5-layer}B and figure~\ref{fig: 1.5-layer}E). We attribute this difference to the zonal asymmetry of baroclinic instability in the intermediate stratification regime. If westward Sverdrup flow is baroclinically unstable then westwards jets will be unstable, too, explaining their disappearance in the 2-layer model. However, we argue that the mechanism for jet creation is the same in both models.\\

To interpret the emergence of the eastward jet, we focus on the transition observed in the 1.5-layer model and propose two complementary perspectives based on turbulence phenomenology in unbounded geometry and western boundary layer dynamics.\\

First, we note that the transition from a gas of isolated vortices to jet-like structures is a common occurrence in stratified two-dimensional turbulence \citep{bouchet2012statistical,venaille2015violent,frishman2017jets}. In the specific case of unbounded isotropic 1.5-layer turbulence, it has long been observed that injecting energy at scales much smaller than the Rossby radius of deformation leads to isolated vortices, sometimes organized into vortex crystals \citep{kukharkin1995quasicrystallization}. In contrast, injecting energy at scales larger than the Rossby radius of deformation results in the formation of large-scale potential vorticity staircases \citep{arbic2003coherent,burgess2022potential}. By definition of potential vorticity, the interfaces of these staircases are associated with sharp jets of width $L_d$: these are the meandering ribbons, which share strong similarities with the eastward jet detaching from the western boundary. A physical interpretation for the emergence of such potential vorticity staircases has been proposed as the most probable outcome of turbulent potential vorticity mixing \citep{venaille_ribbon_2014}.\\

These previous studies suggest that the key parameter governing the transition from a vortex gas to Gulf Stream-like jets is the ratio of the energy injection length scale to the Rossby radius of deformation. In the vortex gas regime, the inertial boundary layer thickness $\delta_I$ sets the maximum scale of the eddies injected from the boundary into the bulk, eventually leading to a vortex gas \citep{miller_gyre_2024}. This suggests that the transition from a vortex gas to ribbon states is governed by $\delta_I/L_d=\sqrt{\xi}$, which is consistent with the transition occurring at $\xi=1$.\\

The presence of $\beta$ and impermeable boundaries in our simulation may change details of this turbulence-driven transition. We note that isolated vortices drift westward at a speed of $\beta L_d^2$, and that the vortex gas regime depends on the ability of these vortices to efficiently interact with the western wall. This interaction is prevented in the eastward parts of the gyres when $U_{sv}>\beta L_d^2$, which again amounts to $\xi > 1$. Thus, the condition for the existence of the vortex gas regime is also consistent with the observed transition at $\xi=1$.\\

So far, we have explained how turbulence may drive a change in the flow pattern within the domain bulk. We now discuss how these patterns may be connected with changes in the western boundary layer dynamics. The emergence of the eastward jet roughly coincides with the stabilization of the western boundary layer (Figure~\ref{fig: instability}), and we hypothesize this stability to give rise to similar recirculation zones as observed in free-slip solutions \citep{ierley1988inertial}. Excitation of different inertial recirculation modes, as outlined by \cite{marshall_zonal_1992}, might also be relevant during the transition. In the asymptotic limits, the 1.5-layer solutions are characterised by a pair of contra-rotating vortices near the western boundary when $\xi \ll 1$ (mean flow not shown, but similar to left panels of figure~\ref{fig: param_space}) and by Fofonoff flow when $\xi \gg 1$, resembling inertial runaway (\cite{sheremet1995analysis}, figure~\ref{fig: 1.5-layer}F). However, when $\xi \sim 1$, both a strong modon and a penetrating jet are present in the outflow region of the western boundary layer (figure~\ref{fig: 1.5-layer}E), and discussing the flow in terms of stationary inertial solutions only does not seem adequate. Although a mechanism setting the pattern of inertial recirculation is left to be identified, these runs are readily compared to the 2-layer simulations. Intense westward parts of inertial recirculation dissapear in the 2-layer runs (\ref{fig: 1.5-layer}B), which is attributed to zonal assymetry of baroclinic instability in surface intensified configurations.\\

In conclusion, the emergence of a strong eastward jet detaching from the western boundary is driven by the dynamics of the 1.5-layer model, while the two-layer dynamics are essential for preventing the formation of intense recirculation and westward jets. This is the central result of this paper.\\

\section{The Zonostrophic Regime ($10 \lesssim \xi$): Freely Decaying Turbulence on Eastward Flow}
\label{sec:zonostrophic-regime}

\begin{figure*}[!htb]
    \centering
    \includegraphics[width = 1\textwidth]{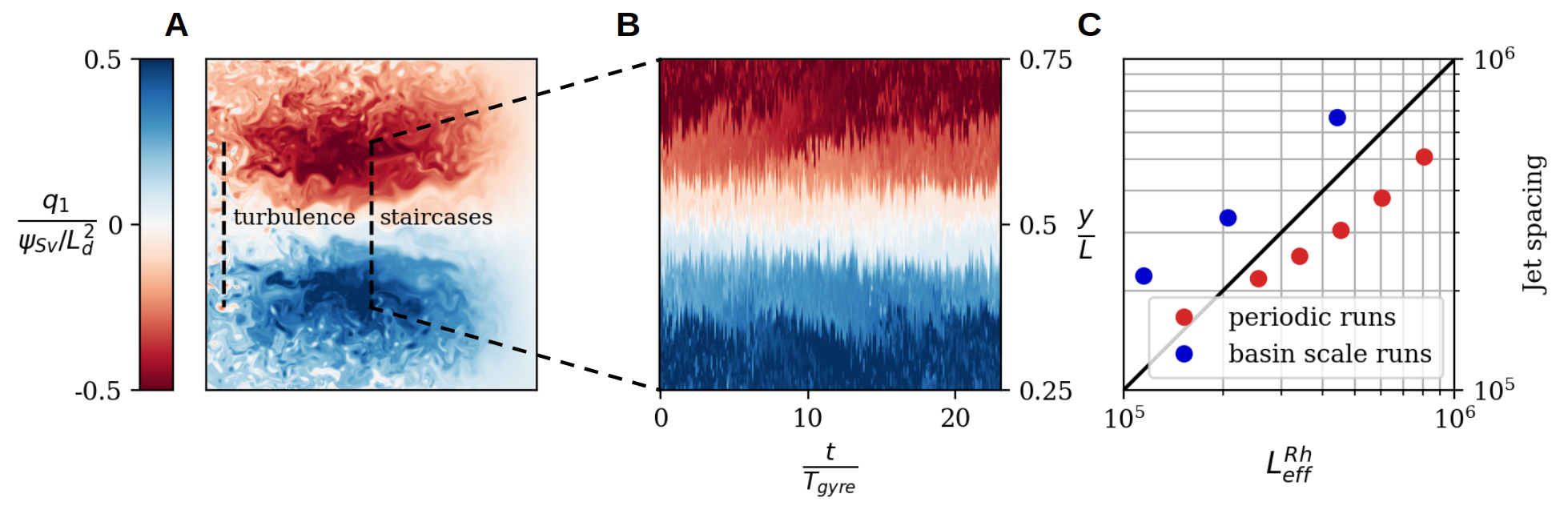}
    \caption{(\textbf{A}) Snapshot of potential vorticity staircases at $\delta = 0.01, \xi = 33$ with (\textbf{B}) Hovmöller diagram taken at $x/L = 1/2$ and $1/4 < y/L < 3/4$. Time is non-dimensionalised by the gyre turnover time $T_{Gyre} = L/U_{Sv}$. (\textbf{C}) Comparison of jet scaling with effective Rhines scale $L^{Rh}_{eff}$. The basin scale runs plotted on this curve are those at $\xi = 10, 32$ and $100$.}
    \label{fig: stairs_basin_scale}
\end{figure*}

If $\xi$ is increased further in the intermediate regime, a loss of western intensification occurs and the system enters a zonostrophic regime (figure~\ref{fig: stairs_basin_scale}). Multiple zonal jets populate regions of eastward flow, and a soup of baroclinic eddies forms in regions of westward flow and close to the western boundary. This regime is likely the same as reported on in \cite{Nadiga_Straub_2019}. In this section, we show that it is best understood as the consequence of strong zonal asymmetry of baroclinic instability.\\

Following the idea that westward flow produces eddies at scales close to $L_d$ it is possible to obtain a scaling for the turbulent velocity scale. If the flow is mostly constrained to the upper layer, we obtain a scaling relation by balancing the energy injection through a Sverdrup interior with dissipation close to a scale $L_d$. Omitting interactions with the lower layer, the upper layer energy balance reads

\begin{equation}
    \int \frac{\boldsymbol{\tau}.\mathbf{u}_1}{H_1} \ dA = \nu \int (\nabla^2\psi)^2 \ dA
\end{equation}
which scales as
\begin{equation}
    \frac{\tau_0 U_{Sv} L^2}{H_1} \sim \frac{\nu U_{eddy}^2 L^2}{L_d^2} 
\end{equation}
and leads to
\begin{equation}
    \frac{U_{eddy}}{U_{Sv}} = \frac{C}{\sqrt{\tilde{\nu}\xi}},
    \label{eq: U_turb}
\end{equation}
where $C$ is a constant. Figure~\ref{fig: global diags}B shows this relation with $C = 0.05$. $U_{eddy}$ is thought to describe the eddies observed in the regions of mean westward flow in the zonostrophic regime, and fits well with the total average speed (figure~\ref{fig: global diags}B). \\

In region of eastward flow, multiple zonal jets form (figure~\ref{fig: jet_transition}, \ref{fig: stairs_basin_scale}A). In the absence of western intensification these jets can no longer be considered as direct extensions of an overshooting western boundary current as in the previous section, neither can they be the product of a purely local inverse cascade as in \cite{berloff_model_2009,berloff_mechanism_2009} because of the stability of the background Sverdrup flow. Here we show that these jets are the result of turbulence, generated through baroclinic instability in the westward flowing regions, which freely decays in a region of stable eastward flow.\\

Turbulence on a $\beta$-plane has a tendency to form zonal jets by creating locally homogenized regions of potential vorticity commonly referred to as staircases \citep{dritschel_multiple_2008} with a spacing of $L^{Rh} = 2\pi\sqrt{U_{turb}/\beta}$ \citep{rhines1975waves}, where $U_{turb}$ is a turbulent velocity scale of the flow. Here, in a 2-layer model with the deeper layer at rest, the surface layer dynamics around an eastward mean flow $U_{m}$ see an effective beta-effect $\beta + U_{m}/L_d^2$ \citep{burgess2022potential}. The jet spacing then scales like
\begin{equation}
    \label{Rh_eff}
    L_{eff}^{Rh} = 2\pi\sqrt{\frac{U_{turb}}{\beta + \frac{U_m}{L_d^2}}} . 
\end{equation}
This formula holds under the assumption of an upper layer surface-intensified eastward flow, which breaks down when this mean flow is baroclinically unstable. Considering stretching by Sverdrup flow, this occurs in gyres at $\xi=1/\delta$. Therefore, when $\delta$ is asymptotically small, there remains a range of criticality parameters $\xi \gg 1$ such that (\ref{Rh_eff}) is valid, with an effective beta term that is dominated by the stretching term induced by eastward Sverdrup flow ($U_{m}/L_d^2 \gg \beta$).\\

We argue that this stretching is responsible for the appearance of multiple eastward jets and the creation of potential vorticity staircases in our simulations at $\xi = 10$, $33$ and $100$. The scaling in equation (\ref{Rh_eff}) is confirmed in figure~\ref{fig: stairs_basin_scale}C. Following the idea of non-local generation of turbulence, $U_{turb}$ was calculated close to the western boundary and the jet scale and $U_m$ in the bulk (figure~\ref{fig: stairs_basin_scale}A, ``turbulence'' and ``staircases'' respectively). To further support the enhanced stretching on an imposed background flow, the scarce data available from the basin-scale runs was complemented with periodic simulations of decaying surface-intensified turbulence with an imposed background flow. Although slightly offset, they also follow the jet spacing given by equation (\ref{Rh_eff}). For further details on the periodic runs, extraction of $U_{turb}$, $U_m$ and the jet scale the reader may consult Appendix~D. \\

Replacing the scaling for $U_{turb}$ by $U_{eddy}$ and $U_{m}$ by $U_{Sv}$ in equation (\ref{Rh_eff}), it is possible to predict the onset of the zonostrophic regime by equating $L_{Rh}^{eff}$ and the size of the region of eastward flow, $L/2$. Although the result depends on a number of constants in front of the scaling relations, simply inserting $U_{eddy}$ as shown in figure~\ref{fig: global diags}B into equation (\ref{Rh_eff}) yields a predicted regime change at $\xi \approx 9$, consistent with numerical experiments. At higher criticality, a precise prediction of the jet scale fails, arguably due to the difference between local and global values of root-mean-square velocity and an incorrect stretching due to the departure from Sverdrup flow at large scales.\\

In summary, the zonostrophic regime is a result of the strong asymmetry of baroclinic instability in the limit of small $\delta$. Sverdrup theory and classic descriptions of western boundary currents no longer match the observed flow, even on a qualitative level. Instead, a clear separation of scales between mean flow heterogeneity and the size of the turbulent structures appears, and the regime is well understood in the framework of unbounded turbulence on an effective $\beta$-plane.\\ 

\section{Discussion and Conclusion}
\label{sec:conclusion}

We found that stratification properties play a crucial role in the emergence of surface-intensified eastward jets in two-layer quasi-geostrophic models without bottom friction. The important nondimensional numbers in this context are the criticality parameter $\xi$, which decreases with the density difference between layers, and $\delta$, the ratio of the depths of the two layers. \\

When the criticality parameter is smaller than one, turbulence takes the form of a western-intensified vortex gas. An energetic eastward jet at the inter-gyre boundary only emerges when $\xi$ is roughly between $1$ and $10$. As the criticality parameter increases further the flow enters a zonostrophic regime, with multiple jets filling the eastward part of the gyres and baroclinic eddies occupying the westward part. Eventually, when $\xi > 1/\delta$, the flow starts to become barotropic.\\

The emergence of the eastward jet detaching from the western boundary when $\xi > 1$ can be understood from two perspectives. First, linear stability analysis indicates that the jet's formation coincides with the stabilization of the inertial western boundary layer, suggesting an interpretation of the westward jet as an overshooting boundary layer. Second, the observed transition from a vortex gas to eastward jets also occurs in unbounded 1.5-layer dynamics when the energy injection length scale exceeds the Rossby radius of deformation $L_d$. Here, we argue that both mechanisms, albeit of different nature, are active in the two-layer model of wind-driven circulation when $\xi \sim 1$.\\

In the 1.5-layer model, the eastward jet emerges similarly to the two-layer case when $\xi$ increases above one. However, this process also leads to westward jets, which are not present in the two-layer model. We showed that these westward jets are disintegrated by baroclinic instability, 
explaining the establishment of the western boundary current extension as a single eastward jet.\\

In this study we investigated values of $\delta$ smaller than the usual oceanic configuration, but the flow dynamics at small $\delta$ remain relevant for understanding the classical configuration. Despite the asymptotically small value of $\delta$, the flow of the reference case shares many properties with the flow obtained with $\delta \sim 0.2$, free-slip boundary conditions and bottom friction \citep{holland_role_1978} (see also appendix~A). The underlying dynamical similarity between the classical flow configuration and the regime at $\delta = 0.01$ becomes apparent in the results of linear stability analysis. Both configurations feature unstable westward Sverdrup flow, but eastward Sverdrup flow and inertial boundary currents are stable. Less extreme surface-intensification will likely shift the exact location of the eastward jet in the parameter space of a more realistic model. If understood as the transitory state between a vortex gas and a zonostrophic regime (figure \ref{fig: jet_transition}), its formation may be altered for example by bottom friction \citep{held1996scaling, gallet2021quantitative} or topographic stretching close to the western boundary \citep{boland2012formation}.\\

A common concern with quasi-geostrophic models of ocean gyres is how sensitive they are to boundary conditions. While free-slip boundary conditions tend to produce more intense eastward jets \citep{haidvogel_boundary_1992, deremble.hogg.ea:on, nasser.madec.ea:sliding}, we showed that jets still form in a no-slip configuration if the stratification is appropriately chosen. 
One reason for this insensitivity to boundary conditions observed here might be the stability of the inertial western boundary layer in the two-layer system (Fig.~\ref{fig: instability}), which is easier to model with free-slip boundary conditions \citep{ierley1991viscous}.\\

The main result of this article is that in a baroclinic model the gyre pattern and the eastward jet are primarily shaped by stratification. This contrasts with the strong dependence of the structure of turbulent baroptropic gyres on dissipative processes \citep{fox2004wind1}. To support the lesser role of dissipation in baroclinic configurations, we note the striking similarity between the free-slip, bottom friction run (figure~\ref{fig:bottom_fric}) and the no-slip, no bottom friction run (figure~\ref{fig: ref_run}), and emphasize that bottom friction is not required to rationalize the emergence of the jet. However, an efficient dissipation mechanism is required to prevent the formation of an energetic vortex gas that may disrupt jet formation (appendix~A). Of course, these energetic flows violate quasi-geostrophic assumptions \citep{scott_small_1998} and are prone to ageostrophic energy sinks \citep{dewar2010topographic,  nikurashin2013routes, bruggemann2015routes}, which, besides bottom friction, will likely become important for dissipating eddies in more realistic scenarios.\\

We conclude that the minimal ingredients for the emergence of a coherent eastward jet extending from western boundary layers in turbulent quasi-geostrophic gyres are: (i) a criticality parameter $\xi$ that is sufficiently large to permit baroclinic instability  ($\xi >1$) but sufficiently small to leave surface-intensified Sverdrup flow intact ($\xi < 1/\delta$), (ii) a layer depth aspect ratio $\delta$ sufficiently small for baroclinic instability to exhibit zonal asymmetry but release energy towards eddies in adjacent areas of westward recirculation, and (iii) an efficient energy sink which suppresses unrealistic eddy-driven flows.\\

A natural extension of the present work is to use these findings to understand the dynamical regimes observed in the world ocean. The first step will be to go beyond the quasi-geostrophic model, which assumes a prescribed, horizontally homogeneous stratification profile. The second step will be to generalize the results obtained from a two-layer model to continuously stratified models. We are currently investigating these two issues using both climatological data and primitive equations model runs.\\

\acknowledgments
 This project has received financial support from the CNRS through the 80 Prime program, and was performed using HPC resources at PSMN Lyon and GENCI HPC (allocation A0150112020).

%
%
\datastatement
Data used in this article were produced with the code qgw (https://doi.org/10.5281/zenodo.13990523). Python notebooks for calculation of potential vorticity homogenization and linear stability analysis are available online (https://doi.org/10.5281/zenodo.14055345). Please contact the corresponding author for access to simulation data.

\clearpage







%



\appendix[A] 
\appendixtitle{Comparison to Realistic Values of $\delta$}
\label{appendixA}

\begin{figure*}[!htb]
    \centering
    \includegraphics[width=0.9\linewidth]{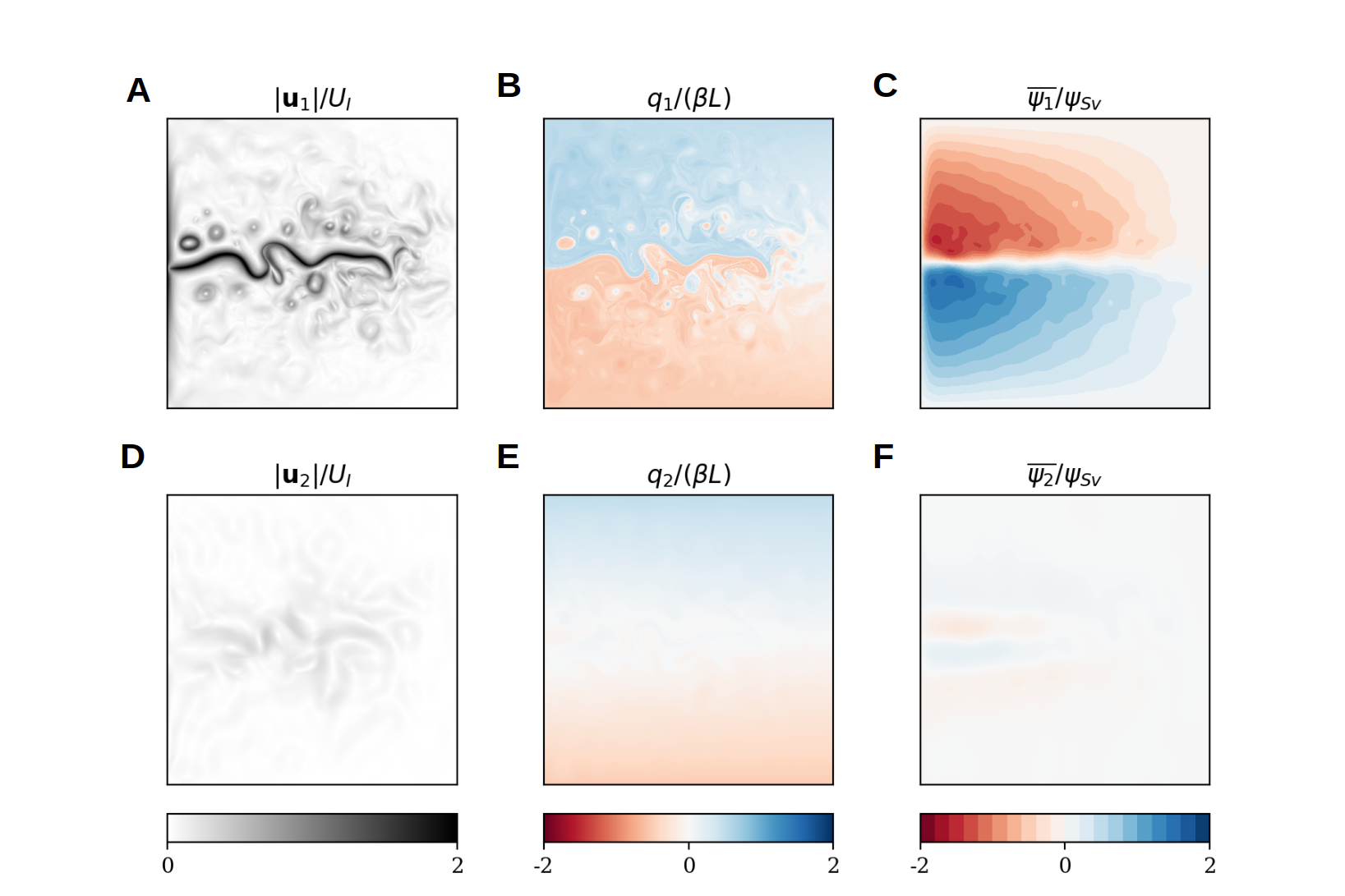}
    \caption{(\textbf{A}, \textbf{D}) Flow speed, (\textbf{B}, \textbf{E}) potential vorticity and (\textbf{C}, \textbf{F}) mean stream function of a simulations at $\delta = 1/6$ ($H_1 = 666$ m, $H_2 = 3334$), $\xi = 1.3$, with free-slip boundary condition and a drag coefficient $r = 3.3\times 10^{-7}$ s$^{-1}$. Note the similarity between this regime and the reference run presented in figure~\ref{fig: ref_run}.}
    \label{fig:bottom_fric}
\end{figure*}

A more traditional approach to modelling energy dissipation in quasi-geostrophic wind-driven gyres is to apply free-slip boundary conditions and model bottom friction via the inclusion of a linear drag term of the form $-r\nabla^2\psi_2$ on the right hand side of the equation for the bottom layer in equation (\ref{dynamical_system2}) \citep{holland_role_1978}. A standard run, at $\delta = 1/6$, is shown in figure~\ref{fig:bottom_fric}, sharing many of the properties listed for the reference run: western and surface intensification, homogenization of potential vorticity inside the gyres and the presence of a strong eastward jet at the interface between the homogenized potential vorticity pools. 

\begin{figure*}[!htb]
    \centering
    \includegraphics[width=0.6\linewidth]{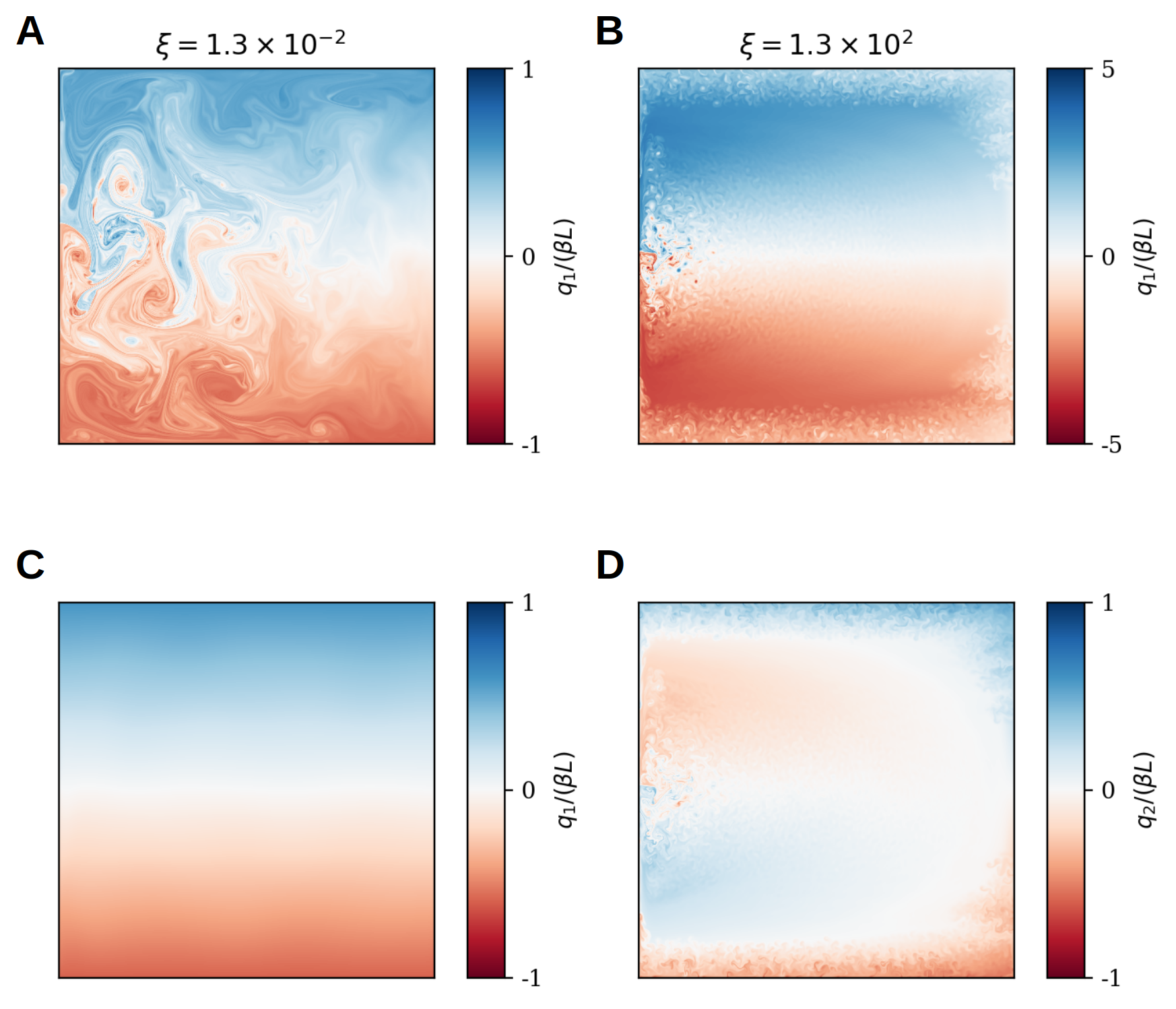}
    
    \caption{Snapshots of upper layer (\textbf{A}, \textbf{B}) and lower layer (\textbf{C}, \textbf{D}) potential vorticity of the free-slip, bottom friction configuration as in figure \ref{fig:bottom_fric} but in the strong $\left(\xi = 1.3\times 10^{-2}\right)$ and weak  $\left(\xi = 1.3\times 10^{2}\right)$ stratification regime. In both regimes the jet disappears and is replaced by energetic vortices close to the western boundary.}
    
    \label{fig: bottom_fric_weak_strong}
\end{figure*}

The jet also disappears in this configuration when entering the regimes of strong and weak stratification regimes in this configuration, as illustrated in figure \ref{fig: bottom_fric_weak_strong}. Strong vortices appear close to the western boundary current as in the no-slip configuration where energy is dissipated by vorticitiy filaments. In the strong stratification regime the flow is completely confined to the upper layer, while in the weak stratification regime the flow becomes completely barotropic. The flow dynamics of these regimes are investigated in \cite{greatbatch2000four}.\\

\begin{figure*}[!htb]
    \centering
    \includegraphics[width=\linewidth]{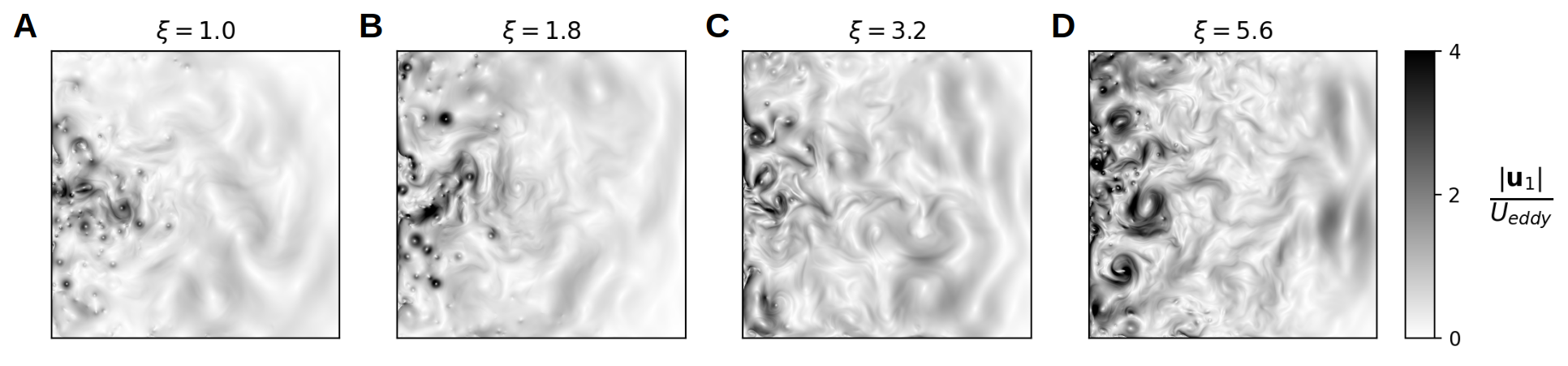}
    \caption{Upper layer flow speed of runs without bottom friction at $\delta = 1/6$ ($H_1 = 666$ m, $H_2 = 3334$ m) for (\textbf{A}) $\xi = 1$, (\textbf{B}) $\xi = 1.8$, (\textbf{C}) $\xi = 3.2$ and (\textbf{D}) $\xi = 5.6$. Note the absence of jets.}
    \label{fig: real_delta_no_jet}
\end{figure*}

If bottom friction is replaced by a no-slip boundary condition as an energy sink, an energetic vortex gas was observed instead of the jet at $\xi = 1$, $\delta = 1/6$  (figure~\ref{fig: real_delta_no_jet}). Jet solutions emerged under three conditions: (i) with the addition of bottom friction, (ii) with an increase in viscosity, or (iii) with a decrease in the layer depth aspect ratio. Its reappearance at lower $\delta = 0.01$, presented as the reference run in the main text, may be understood as the consequence of the stability island of the western boundary layer becoming larger as $\delta$ is decreased (figure~\ref{fig: instability}). An inertial overshoot of the western boundary layer remains possible at smaller values of $L_d$, where smaller vortices dissipate energy more effectively (equation (\ref{eq: U_turb})). These vortices thus become less energetic, and the jet may support their presence. \\

\appendix[B] 
\appendixtitle{Details on the Regime of Weak Stratification}
\label{appendixB}

\subsection*{Potential Vorticity Homogenization Theory}

Calculations on potential vorticity homogenization in 2-layer oceanic gyres can be found in classical textbooks \citep{vallis2017atmospheric}. A simple extension including unequal layer depths is given here, with a focus on the mechanism responsible for the onset of homogenization in the lower layer. We write the quasi-geostrophic equations on a modal basis, neglecting time derivatives and relative vorticity contributions, but including diffusion of potential vorticity instead of relative vorticity.

\begin{align}
    \beta\frac{\partial \psi_{bt}}{\partial x} &= \frac{\nabla\times \tau}{H_1 + H_2}\label{equ: bc_bt1} \\
    \beta\frac{\partial \psi_{bc}}{\partial x}&=  \frac{1}{L_d^2}J(\psi_{bt}, \psi_{bc}) + \frac{\nabla\times \tau}{H_1}  + \nu \frac{\nabla^2\psi_{bc}}{L_d^2}  \label{equ: bc_bt2}\\
    \psi_{bt} &= \frac{H_1 \psi_1 + H_2 \psi_2}{H_1 + H_2} \\
    \psi_{bc} &= \psi_1 - \psi_2
\end{align}

When $L_d$ is large, the advective term for the baroclinic mode can be neglected and all transport happens in the upper layer. A change in regime occurs when the advective term of the baroclinic mode starts to be of the same order of magnitude as the $\beta$-term. Using Sverdrup scalings for $\psi_{bc/bt}$ as obtained from equation (\ref{equ: bc_bt1},\ref{equ: bc_bt2}), this occurs when

\begin{equation}
    1 \sim \frac{\frac{1}{L_d^2}J(\psi_{bt}, \psi_{bc})}{\beta\frac{\partial \psi_{bc}}{\partial x}} \sim \frac{U_{Sv} H_1}{\beta L_d^2 (H_1 + H_2)} \sim \delta\xi.
\end{equation}

Baroclinic instability is not required for this change to occur, it is only the baroclinic/barotropic advection that alters the dynamical balance of the Sverdrup flow. However, baroclinic instability will always be present at the onset of inertial recirculation in surface-intensified flows, its sole effect being represented by the presence of weak diffusion of potential vorticity. This diffusion sets the final flow by homogenizing potential vorticity in the lower layer. The solution is then given by matching the contours at the center latitude $y= L/2$ and seperating regions of blocked and closed geostrophic contours. The solution is written in terms of

\begin{align}
    \overline{\psi} &= \frac{2\pi\tau_0}{\beta (H_1 + H_2)}\sin\left(\frac{2\pi y}{L}\right)\left(1-\frac{x}{L}\right) \\
    \overline{q} &= \beta y + \frac{\overline{\psi}}{L_d^2}
\end{align}

Geostrophic contours are blocked in the northern gyre if $\overline{q} > \ \beta L/2$ and in the southern gyre if $\overline{q} < \ \beta L/2$. The upper layer solution is then given by

\begin{equation}
    \psi_1 = \frac{(H_1 + H_2)\overline{\psi}}{H_1}.
\end{equation}
Else, for a closed contour, homogenization of potential vorticity in the lower layer leads to
\begin{equation}
    \label{closed}
    \psi_1 = \overline{\psi} - \frac{H_1}{H_2}\beta L_d^2\left(y - \frac{L}{2}\right).
\end{equation}
The lower layer $\psi_2 = \left((H_1 + H_2)\overline{\psi} - H_1 \psi_1\right)/H_2$ carries the remaining transport to satisfy the barotropic Sverdrup balance. The lines in figure~\ref{fig: global diags}A are computed numerically from this solution. Details on this numerical calculation may be found in the code repository linked in the Data availability statement.\\

\subsection{Observed Flows at $\delta = 0.01$}

\begin{figure*}[!htb]
    \centering
    \includegraphics[width = 0.6\textwidth]{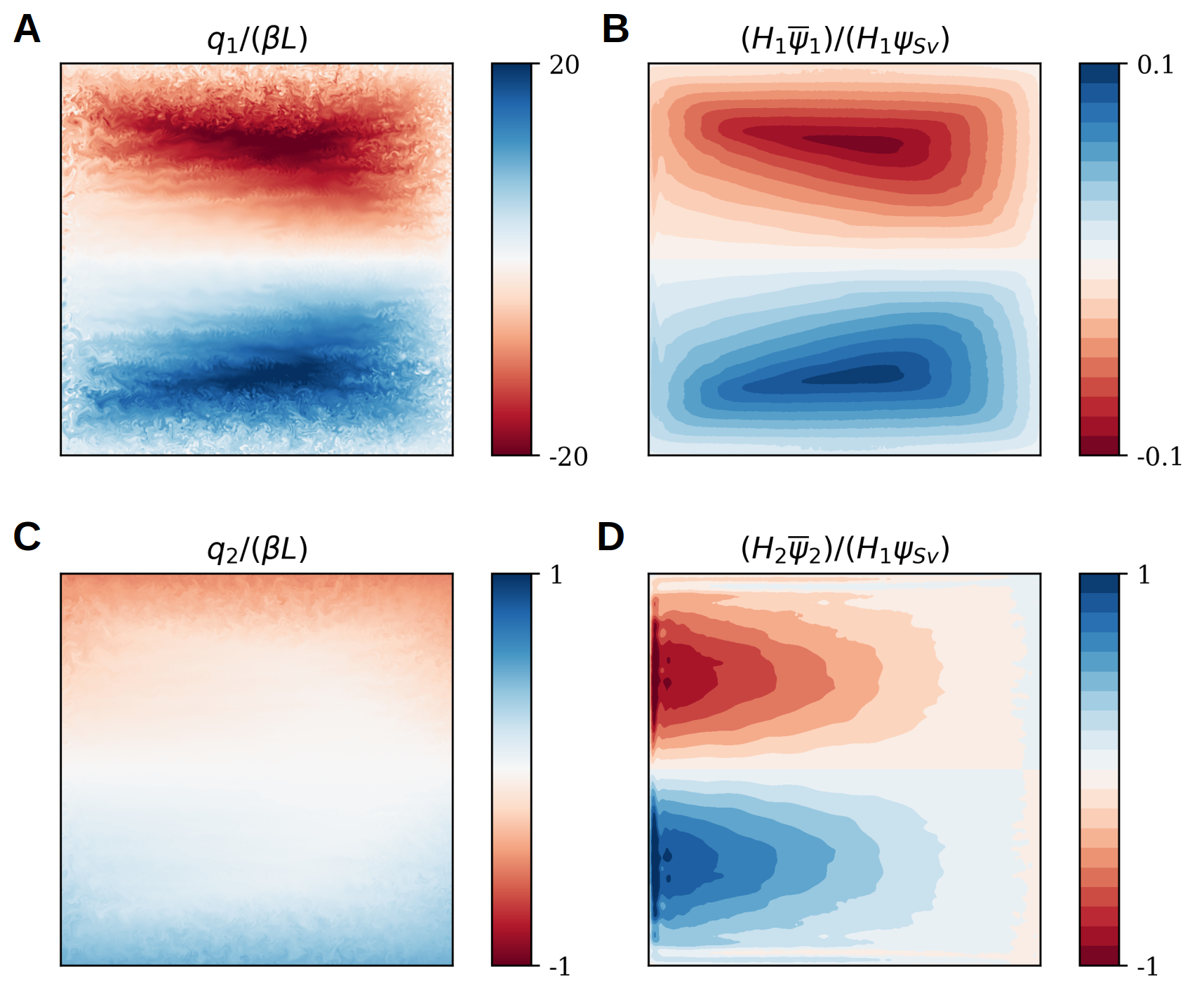}
    \caption{(\textbf{A},\textbf{C}) Potential vorticity snapshots and (\textbf{B},\textbf{D}) mean transport of the weak stratification regime low aspect ratio ($\xi = 10^3$, $\delta = 0.01$). Baroclinic activity remains strong even though a Sverdrup flow starts to grow in the lower layer.}
    \label{fig: weak_strat}
\end{figure*}

At equal layer depths a complete barotropization occurs when $\delta\xi > 1$ (figure~\ref{fig: global diags}). When $\delta$ is small, however, both the mean flow and the turbulent flow features retain a strong baroclinic signature (figure~\ref{fig: weak_strat}). In the lower layer a Sverdrup Flow starts to grow and potential vorticity homogenizes. The upper layer appears to stay in the zonostrophic regime, with strong zonal asymmetry of the turbulent flow features and a complete loss of western intensification. We discuss here the asympotic nature of this flow as $\xi \rightarrow \infty$. From equation \ref{closed} it can be shown that even in the homogenized region the baroclinic zonal shear leads to a criticality of

\begin{align}
    \frac{u_{bc}}{\beta L_d^2} = \frac{1}{\delta},
\end{align}

rationalising the strong activity of baroclinic instability in the weakly stratified regime when $\delta$ is small. Nonetheless, a complete barotropization is expected at larger $\xi$ than those explored here (figure~\ref{fig: global diags}A), although viscosity might prevent barotropization of turbulence by direct dissipation in the upper layer (equation \ref{eq: U_turb}). Remarkably, potential vorticity homogenization theory works indifferent to the complex dynamics in the upper layer. \\

\appendix[C]
\appendixtitle{Instability Calculations}
\label{appendixD}
Linearising the QG equations around a base state with upper layer velocity only, we obtain:

\begin{align}
    \frac{\partial q_1}{\partial t} + J(\psi_1,Q_1) + J(\Psi_1, q_1) = \nu \nabla^4\psi_1 \\ \frac{\partial q_2}{\partial t} + J(\psi_2,Q_2) =  \nu \nabla^4\psi_2 
\end{align}

Here $q_{1/2}, \psi_{1/2}$ denote the perturbation fields but are defined as in the main text. The base state is given by

\begin{equation}
    Q_1 = \nabla^2\Psi_1 - \frac{1-\delta}{L_d^2}\Psi_1 + \beta y\ , \ Q_2 = \frac{\delta}{L_d^2}\Psi_1 + \beta y.
\end{equation}

For the linear stability analysis, three base states were defined. For the interior flow, the base state was given by $\Psi_1 = \pm U_{Sv} y$, where the $\pm$ stands for eastward or westward flow respectively. In these calculations, $\nu$ was set to zero. The problem is governed by two non-dimensional numbers, $\xi$ and $\delta$.\\

The instability analysis of the western boundary layer was carried out by assuming that the problem is invariant in the y-direction. The base profile is defined as in equation \ref{eq: base profile} and meant to resemble a double-deck boundary layer structure with an inertial thickness of $\delta_I$ and a viscous sublayer $\delta_P$. As in the simulations, the sublayer thickness was set to $\delta_P \approx 3$ km and the inertial layer thickness to $\delta_I \approx 72$ km. The explicit dependence on $\delta_I$ and $\delta_P$ is not explored further.\\

The instability problem for interior flow is decomposed on Fourier modes and analytically solvable, but the expressions are cumbersome and not very insightful. They can also be found in \cite{pedlosky1987geophysical} and will therefore not be reproduced explicitly, and the reader is referred to figures \ref{fig: instability} for results. For the western boundary current, decomposition in $y$ is carried out on a Fourier basis, too, and the eigenvectors in $x$ are solved for numerically using 4th order finite-element discretisation of the derivatives. Details on this calculation may be found in the code repository linked in the Data availability statement. Again, the results are shown in the main text.\\ 

\appendix[D] 
\appendixtitle{Details on the Zonostrophic Regime}
\label{appendixC}

\begin{figure*}[!htb]
    \centering
    \includegraphics[width = \textwidth]{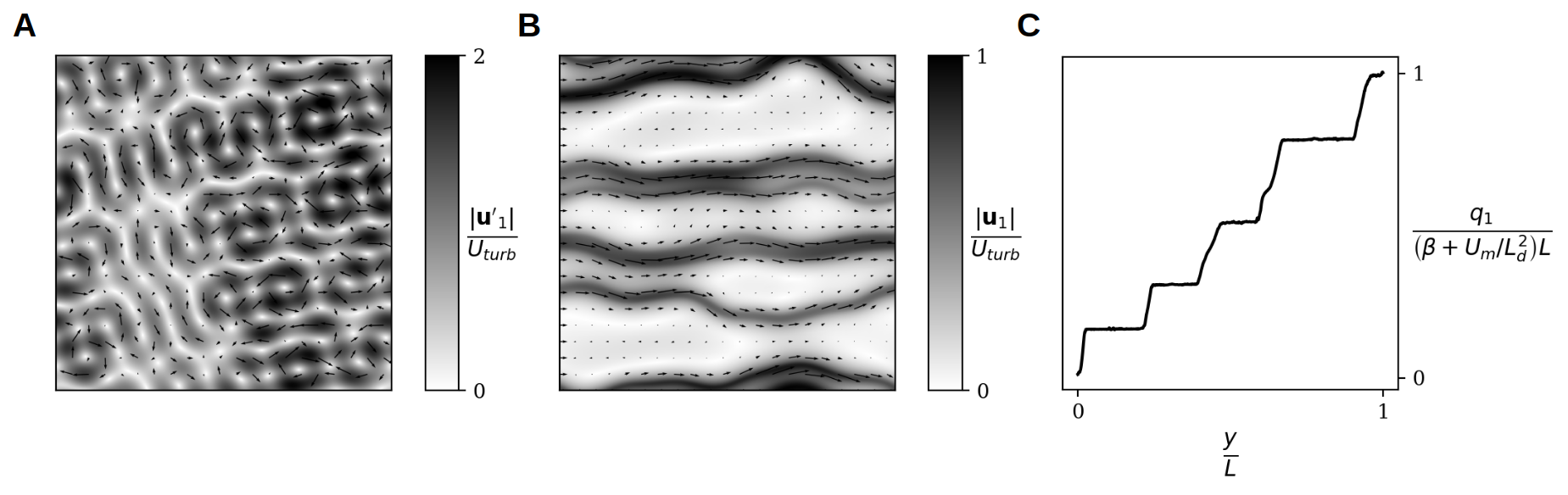}
    \caption{(\textbf{A}) Initial velocity perturbation, (\textbf{B}) final velocity fields and (\textbf{C}) meridional profile of potential vorticity for the periodic staircase run at injection velocity $U_{turb} = 3.2(\beta L_d^2 + U_{m})$.}
    \label{fig: jets}
\end{figure*}

In order to compare jet spacing in the zonostrophic regime with $L_{eff}^{Rh}$, we measured $U_{turb}$ as the average fluctuation speed along the meridional line $L/4 < y < 3L/4$, $x = 3\delta_I$ and $U_{m}$ was measured as the mean transport across the line $L/4 < y < 3L/4$, $x = L/2$ divided by $L/2$ (figure~\ref{fig: stairs_basin_scale}A, lines entitled ``turbulence'' and ``staircases'' respectively). In order to determine the number of jets, we calculate the histogram of the potential vorticity along the Hovmöller diagram at mid-basin (figure~\ref{fig: stairs_basin_scale}B). Finally, to obtain the jet spacing, the meridional size over which the histogram was obtained is then divided by the number of distinct peaks that exceed the standard deviation of the histogram, corresponding to the number of distinct homogenized regions. \\

We performed additional runs with periodic boundary conditions to complement the basin-scale runs. An example simulation can be seen in figure~\ref{fig: jets}, where the velocity perturbation is denoted as $\textbf{u}'$. The initial flow fields are given by

\begin{align}
    \psi_1(t = 0) = \psi'_1 - U_{m}y &\ q_1(t = 0) = q'_1 + \frac{1 - \delta}{L_d^2}U_{m}y \\
    \psi_2(t = 0) = 0 &\ q_2(t = 0) = -\frac{\delta}{L_d^2}U_{m}y 
\end{align}

with $q_1' = \nabla^2\psi_1' - (1-\delta)/L_d^2(\psi_2'-\psi_1') + \beta y$. The stream function was initialised with a randomly generated signal centered around the wavenumber $1/L_d$, meaning to represent turbulence generated close to the western boundary due to baroclinic instability, and then left to freely decay. In all these simulations the deformation radius was fixed at $L_d = 40.7$ km, $U_{m} = 0.2$ m/s, $\delta = 0.01$ and the domain size was set to $L = 1523$ km. It was also necessary to decrease viscosity to $\nu = 0.1$ m$^2$/s in order to obtain final states with clear staircases. This required a numerical resultion of $2048x2048$ grid points, and all simulations were run until mixing of potential vorticity transferred the complete perturbuation energy into zonal jets. All other parameters are the same as in the surface-intensified basin-scale simulations (table \ref{t1}). \\

To show the dependence of the jet spacing on the initial perturbation, we carried out five periodic simulations varying the turbulent velocity scale $U_{turb}$. It is defined here as 

\begin{equation}
    U_{turb} = \sqrt{\frac{1}{L^2}\int |\nabla\psi_1'|^2 \ dA}
\end{equation}

at the time of initiation, and was varied between $\beta L_d^2 + U_{m}$ and $10(\beta L_d^2 + U_{m})$. To determine the jet spacing the histogram of potential vorticity is taken of a snapshot of the final state over the entire domain, and then the same procedure as for the basin-scale simulations is followed to extract the jet scale.\\

Figure~\ref{fig: stairs_basin_scale}B also shows the slow polewards displacement of the staircases of potential vorticity in the basin-scale runs. \cite{Nadiga_Straub_2019} suggested that the displacement speed is correlated with $U_{Sv}$, however in our simulations it seemed to match better with the maximum Rossby wave speed, $\beta L_d^2$.\\


\bibliographystyle{ametsocV6}
\bibliography{references}

\end{document}